\documentclass[mathpazo,a4paper]{cicp}

\usepackage{amssymb,amsmath,algorithm,algorithmic}
\usepackage{color}
\definecolor{lightblue}{rgb}{0.22,0.45,0.70}

\usepackage{natbib}

\numberwithin{equation}{section}
\numberwithin{table}{section}
\numberwithin{figure}{section}

\usepackage[colorlinks=true,breaklinks=true,linkcolor=lightblue,citecolor=lightblue]{hyperref}

\newcommand{\mbs}[1]{\boldsymbol{#1}}
\newcommand{\mbf}[1]{\mathbf{#1}}

    \newcommand\bC{{\mbf{C}}}
\newcommand\bD{{\mbf{D}}}    \newcommand\bF{{\mbf{F}}}
  \newcommand\bH{{\mbf{H}}}  \newcommand\bI{{\mbf{I}}}

\newcommand\bP{{\mbf{P}}}    \newcommand\bR{{\mathbb{R}}}

  \newcommand\bb{{\mbs{b}}}

  \newcommand\bn{{\mbs{\nu}}}  
    
  \newcommand\bt{{\mbs{t}}}  \newcommand\bu{{\mbs{u}}}
\newcommand\bv{{\mbs{v}}}  \newcommand\bx{{\mbs{x}}}

\newcommand{\cS}{\mathcal{S}}
\newcommand{\cG}{\mathcal{G}}
\newcommand{\cH}{\mathcal{H}}
\newcommand{\cR}{\mathcal{R}}
\newcommand{\cT}{\mathcal{T}}
\newcommand\fo{{\mbs{f}_0}}
\newcommand\so{{\mbs{s}_0}}
\newcommand\no{{\mbs{n}_0}}
\newcommand\ko{{\mbs{k}_0}}

\newcommand\btau{{\boldsymbol{\tau}}}
\newcommand\bzeta{{\boldsymbol{\zeta}}}
\newcommand\cero{\boldsymbol{0}}
\newcommand\bPi{{\mbf{\Pi}}}
\newcommand\tr{{\rm tr}\,}

\allowdisplaybreaks

\newcommand{\cred}{}

\begin{document}

\title{Modelling thermo-electro-mechanical effects \\in orthotropic \cred{cardiac} tissue}
\author[R. Ruiz-Baier et.~al.]{Ricardo Ruiz-Baier\affil{1,2}\comma\corrauth,
 Alessio Gizzi\affil{3}, Alessandro Loppini\affil{3}, \\
 Christian Cherubini\affil{3,4}, and Simonetta Filippi\affil{3,4}}
\address{\affilnum{1}Mathematical Institute,
University of Oxford,  Woodstock Road, Oxford OX2 6GG, United Kingdom\\
\affilnum{2}Laboratory of Mathematical Modelling, Institute of Personalised Medicine, Sechenov University, 
Moscow, Russian Federation\\
\affilnum{3}Nonlinear Physics and Mathematical Modeling, Department of Engineering, University Campus Bio-Medico of Rome, Via A. del Portillo 21, 00128, Rome, Italy\\
\affilnum{4}International Center for Relativistic Astrophysics (ICRA), and ICRANet, Piazza delle Repubblica 10, I-65122 Pescara, Italy}
\emails{{\tt ruizbaier@maths.ox.ac.uk} (R. Ruiz-Baier), {\tt a.gizzi@unicampus.it} (A. Gizzi), {\tt a.loppini@unicampus.it} (A. Loppini), 
{\tt c.cherubini@unicampus.it} (C. Cherubini), {\tt s.filippi@unicampus.it} (S. Filippi)}

\begin{abstract}
In this paper we introduce a new mathematical model for the active contraction of cardiac muscle, \cred{featuring} different thermo-electric and nonlinear conductivity properties. The passive hyperelastic response of the tissue is described by an orthotropic exponential model, whereas the ionic activity dictates active contraction incorporated through the concept of orthotropic active strain. We use a fully incompressible formulation, and the generated strain modifies directly the conductivity mechanisms in the medium through the pull-back transformation.  We also investigate the influence of thermo-electric effects in the onset of multiphysics emergent spatiotemporal dynamics, using nonlinear diffusion. It turns out that these ingredients have a key role in reproducing pathological \cred{chaotic dynamics} such as ventricular fibrillation during inflammatory events, for instance. The specific structure of the governing equations \cred{suggests to cast the problem in mixed-primal form and we write it in terms of Kirchhoff stress, displacements, solid pressure, electric potential, activation generation, and ionic variables. We also} propose  a new mixed-primal finite element method for its numerical approximation, and we use it to explore the properties of the model and to assess the importance of coupling terms, by means of a few computational experiments in 3D.
\end{abstract}

\maketitle

\section{Introduction}
Temperature variations may have a direct impact on many of the
fundamental mechanisms in the cardiac function
\citep{wilson:2011}. Substantial differences have been reported in the
conduction velocity and spiral drift of chaotic electric potential
propagation in a number of modelling and computationally-oriented  
studies \citep{filippi:2014},
and several experimental tests confirm that this is the case not only
for cardiac tissue, but for other excitable systems
\citep{gizzi:2010,kienast:2017}. The phenomenon is however not
restricted to electrochemical interactions, but it also might affect 
mechanical properties \citep{hill:1938,lawton:1954,sugi:1998}. 
\cred{Indeed, c}ardiac muscle is quite sensitive to
mechanical stimulation and deformation patterns
can be very susceptive to external agents such as temperature. For
instance, enhanced tissue heterogeneities can be observed when the
medium is exposed to altered thermal states, and in turn these can
give rise to irregular mechano-chemical dynamics. A few examples 
that relate to experimental observations from epicardial and endocardial 
activity on canine right ventricles at different temperatures, as well as 
tachycardia and other fibrillation mechanisms \cred{occurring due to   
thermal unbalance}, can be found in e.g. \cite{filippi:2014}. These scenarios can be related to extreme 
conditions encountered during heat strokes and sports-induced fatigue (easily reaching 41$^\circ$C),
 and localisation of other thermal sources such as ablation devices; 
but also 
to surgery or therapeutical procedures (in open-chest surgery tissues 
might be exposed to cold air in the operating theatre at 25$^\circ$C), or due to 
extended periods of exposure to even lower temperatures that can occur during  
 shipwrecks or avalanches. It is not striking that temperature effects 
might affect the behaviour of normal electromechanical heart activity. However 
the precise form that these mechanisms manifest themselves is not at all 
obvious. This is, in part, a consequence of the nonlinear character of the thermo-electro-mechanical coupling. For instance, one can show that localised thermal gradients might destabilise the expected propagation of the electric wave, as well as change the mechanical behaviour of anisotropic contraction. 
Our goal is to 
investigate the role of the aforementioned effects in the development
and sustainability of cardiac arrhythmias. These complex emerging
phenomena originate from multifactorial and multiphysical interactions
\citep{weiss:2014}, and they are responsible for a large number of
cases of pathological dysfunction and casualties.  The model we
propose here has potential therefore in the investigation of mechanisms
provoking such complex dynamics, 
in particular those arising during atrial and ventricular fibrillation.

\cred{Even if computational models for the 
electromechanics of the heart are increasingly complex and account for
many multiphysics and multiscale effects (see
e.g. \citealp{colli:2016,trayanova:2011a,quarteroni:2017}), we are only aware of
one recent study \citep{collet:2017} that addresses similar questions
to the ones analysed here. However that study is restricted to 
one-dimensional domains, it uses the
two-variable model from \cite{nash:2000}, and it assumes an active stress
approach for a simplified neo-Hookean material 
in the absence of an explicit stretch state.}
Our phenomenological framework also uses a temperature-based
two-variable model, but in contrast, it additionally 
includes a nonlinear conductivity representing a \cred{generalised
diffusion mechanism intrinsic to porous-medium 
electrophysiology \citep{hurtado:2016}}. We postulate then an extended model that also
accounts for active deformation of the tissue, where the specific form
of the electromechanical coupling is dictated by an adaptation of the orthotropic
active strain framework proposed in \cite{rossi:2014}. 

We have structured the contents of this paper in the following 
manner. Section~\ref{sec:model} \cred{discusses a} combination of phenomenological and physiological coupled models from thermo-electric and thermo-mechanic dynamics being local (potentially sub-cellular in a physiological model), tissue, and organ-scale levels. We introduce in Section~\ref{sec:fe} 
a new mixed-primal finite element scheme for the solution of the set of 
governing equations (in particular using the Kirchhoff stress as additional unknown), where we provide also some details about its computational realisation. 
All of our numerical tests are collected in Section~\ref{sec:results}, including conduction velocity assessment, and a few simulations regarding normal and arrhythmic dynamics in simplified 3D domains. We then 
conclude in Section~\ref{sec:concl} with \cred{a summary and a discussion on the limitations 
and envisaged extensions} of this study.

\section{A new model for thermo-electric active strain}\label{sec:model}
In this section we provide an abridged derivation of the set of partial 
differential equations describing the multiscale coupling between electric, 
thermal, mechanical, and ionic processes; which are, in principle, valid for general 
excitable and deformable media. 

\subsection{Muscle contraction via the active strain approach}
Let $\Omega\subset\cred{\bR^3}$ denote a deformable body with
piecewise smooth boundary $\partial\Omega$, regarded in its reference
configuration, and denoted by $\bn$,  the outward unit normal vector on
$\partial\Omega$. The kinematical description of finite deformations
regarded on a time interval $t\in (0,t_{\text{final}}]$ is made
precise as follows.  A material point in $\Omega$ is denoted by $\bx$,
whereas $\cred{\bx_t - \bx=} \bu(t):\Omega\to\cred{\bR^3}$ will denote the displacement field
\cred{characterising} its new position $\bx_t$ within the body $\Omega_t$ in the
current, deformed configuration.  The tensor \cred{$\bF:=\bI+\nabla\bu$} is the
gradient (applied with respect to the fixed material coordinates) of
the deformation map; its Jacobian determinant, denoted by $J=\det\bF$,
measures the solid volume change during the deformation; and
$\bC=\bF^{\tt t}\bF$ is the right Cauchy-Green deformation tensor on
which all strain measures will be based (here the superscript $()^{\tt t}$ denotes 
the transpose operator). The first isotropic
invariant {controlling} deviatoric effects is $I_{1}(\bC)=\tr\bC$, and for
generic unitary vectors $\fo,\so$, the scalars
$I_{4,f}(\bC)=\fo\cdot(\bC\fo)$, $I_{8,fs} (\bC)=\fo\cdot(\bC\so)$ are
direction-dependent pseudo-invariants of $\bC$ measuring fibre-aligned
stretch (see e.g.~\citealp{spencer:1980}).  As usual, $\bI$ denotes the
\cred{3$\times$3} identity matrix. In the remainder of the presentation we
will restrict all space differential operators to the material coordinates.

Next we recall the active strain model for ventricular
electromechanics as introduced in \cite{nobile:2012}. There, the
contraction of the tissue results from activation mechanisms governed
by internal variables and incorporated into the finite elasticity
context using a virtual multiplicative decomposition 
of the deformation gradient into
a passive (purely elastic) and an active part $\bF =
\bF_{\!E}\bF_{\!A}$, defined in \cred{general, triaxial} form 
\begin{equation}\label{eq:FA}
\cred{\bF_{\!A} =\bI+\gamma_f\fo(\bx)\otimes\fo(\bx)+\gamma_s\so(\bx)
\otimes\so(\bx)+\gamma_n\no(\bx)\otimes\no(\bx).}
\end{equation}
The 
coefficients $\gamma_i$, with $i=f,s,n$, are smooth scalar functions encoding the
macroscopic stretch in specific directions, whose precise definition
will be postponed to Section~\ref{sec:activation}.  The inelastic contribution to the deformation 
modifies the \cred{size of} the cardiac fibres, and then compatibility 
of the motion is restored through an elastic deformation accommodating 
the active strain distortion. The triplet 
$(\fo(\bx),\so(\bx),\no(\bx))$ represents a coordinate system pointing
in the local direction of cardiac fibres, transversal sheetlet compound,
and normal cross-fibre direction \cred{$\no(\bx) = \fo(\bx)\times\so(\bx)$.}

\cred{Constitutive} relations defining the material
properties and underlying microstructure of the myocardial tissue will follow the orthotropic model proposed in
\cite{holzapfel:2009}, for which the strain energy function and the
first Piola-Kirchhoff stress tensor (after applying the active strain
decomposition) read respectively 
\begin{equation}\label{piolatensor}
\Psi(\bF_E) = \frac{a}{2b}e^{b(I^E_{1}-\cred{3})}
+\dfrac{a_{fs}}{2b_{fs}}\bigl[e^{b_{fs}(I_{8,fs}^E)^{2}}-1\bigr]+ 
\sum_{i\in\{ f,s\} } \dfrac{a_{i}}{2b_{i}}\bigl[e^{b_{i}\cred{((I^E_{4,i}-1)_+)^{2}}}-1\bigr], \quad 
\bP=\frac{\partial\Psi}{\partial\bF}-pJ\bF^{-{\tt t}},
\end{equation}
where $a,b,a_i,b_i$ with $i\in\{f,s,fs\}$ are material parameters, 
$p$ denotes the solid hydrostatic pressure, and \cred{have used the notation 
 $(x)_+:=\max\{x,0\}$}. Switching off the anisotropic 
contributions \cred{acting on $\so$ and $\no$ (but not the shear term)} 
under compression ensures that the associated terms in the 
strain energy function (in both the pure passive and active-strain formulations) 
are strongly elliptic \citep{pezzuto:2014} (these will be the terms 
appearing on the second diagonal block of the weak formulation from Section~\ref{sec:fe}, 
the block corresponding to displacements), however 
the overall problem will remain of a saddle-point structure. 

The modified elastic invariants $I_i^E$ are functions of
the coefficients $\gamma_i$ and the invariant and pseudo invariants as
follows
\begin{align*}
I_{1}^{E} & = \cred{\biggl[1-\frac{\gamma_n(\gamma_{n}+2)}{(\gamma_{n}+1)^2}\biggr]I_{1} 
+\biggl[\frac{\gamma_{n}(\gamma_{n}+2)}{(\gamma_{n}+1)^2}
-\frac{\gamma_{f}(\gamma_{f}+2)}{(\gamma_{f}+1)^2}\biggr]I_{4,f} 
+\left[\frac{\gamma_{n}(\gamma_{n}+2)}{(\gamma_{n}+1)^2}
-\frac{\gamma_{s}(\gamma_{s}+2)}{(\gamma_{s}+1)^2}\right]I_{4,s}},\\
I_{4,f}^E & =\dfrac{I_{4,f}}{\left(\gamma_{f}+1\right)^{2}},\quad  
I_{4,s}^E  =\dfrac{I_{4,s}}{\left(\gamma_{s}+1\right)^{2}},\quad  
I_{8,fs}^E  =\dfrac{I_{8,fs}}{\left(\gamma_{f}+1\right)\left(\gamma_{s}+1\right)}.
\end{align*}
Accordingly, the active strain and consequently the force associated to the 
active part of the total stress, will receive contributions acting 
distinctively on each direction $\fo(\bx),\so(\bx),\no(\bx)$.  

The balance of linear momentum together with the 
incompressibility constraint are 
written, when posed in the inertial reference frame and under 
pseudo-static mechanical equilibrium, in the following way
\begin{subequations}\label{eq:mech}
\begin{align}
- \boldsymbol{\nabla} \cdot \bP & = \rho_0 \bb & \text{in } \Omega\times(0,t_{\mathrm{final}}], \label{eq:momentum}\\ 
\rho J - \rho_0 & = 0 & \text{in } \Omega\times(0,t_{\mathrm{final}}], \label{eq:mass}
\end{align}
\end{subequations}
where $\rho_0,\rho$ are the reference and current 
medium density, and $\bb$ is a vector of body loads. \cred{Furthermore,} the 
balance of angular momentum translates into the condition of 
symmetry of the Kirchhoff stress tensor $\bPi= \bP\bF^{\tt t}$, which is in turn 
encoded into the momentum and constitutive relations \eqref{eq:momentum}, \eqref{piolatensor}, \cred{and \eqref{eq:FA}}. 

\cred{Following the notation in \cite{chavan07}, the contribution to stress that does not include pressure explicitly is denoted as 
$\cG(\bu) := \frac{\partial\Psi}{\partial\bF}\bF^{\tt t}$, and therefore we have the constitutive relation} 
\begin{equation}\label{eq:Pi}
\bPi = \cG(\bu) - p J\bI.
\end{equation}

\subsection{A modified Karma model for cardiac action potential}
Let us denote by $I_{\rm ext}$ a spatio-temporal external electrical stimulus
applied to the medium. On the undeformed configuration we proceed to
write the following monodomain equations describing the transmembrane
potential propagation and the dynamics of slow recovery currents
according to a specific temperature $T$:
\begin{subequations} \label{eq:genKarma}
\begin{align}
  \label{eq:genKarmaV}
  \frac{\partial v}{\partial t} -  
  \nabla \cdot {\left[ \bD(v,\bF)\, \nabla v \right] }
  &=  \dfrac{f(v,n)}{\tau_v(T)} + I_{\rm ext} & \text{in } \Omega\times(0,t_{\mathrm{final}}],\\
  \label{eq:genKarmaN}
  \frac{dn}{dt} & = \dfrac{g(v,n)}{\tau_n (T)} & \text{in } \Omega\times(0,t_{\mathrm{final}}],
\end{align}
\end{subequations}
where the unknowns are the transmembrane potential, $v$, and 
the recovery variable, $n$. 
This reaction-diffusion system is endowed with the following
specifications, taking the membrane model proposed in
\cite{karma:1994}, and adapting it to include
thermo-electric effects following the development in \cite{gizzi:2017}
\begin{subequations}\label{eq:reactions}
\begin{align}
	f(v,n) &:= -v + [v^* - \cS(n)] [1 - \tanh(v-v_\star)] \frac{v^2}{2},
	\\
	g(v,n) &:= \cR(n) \cH(v-v_n) - [1-\cH(v-v_n)] n,
	\\
	\label{eq:RnDn}
	\cR(n) &:= \dfrac{1-(1-e^{-L}) n}{1-e^{-L}}
	,\qquad\quad
	\cS(n) := n^M,\\
	\label{eq:tauT}
	\tau_v (T) &:= \dfrac{\tau_v^0}{1 + \beta(T-T_0)}
	,\quad
	\tau_n (T) := \tau_n^0\,\cred{Q_{10}(T)}.
\end{align}
\end{subequations}
\cred{As in the original phenomenological model from \cite{karma:1994},
here $\cH(x)$ stands for the Heaviside step function, i.e. $\cH(x)=0$
for $x\le0$ and $\cH(x)=1$ for $x>0$. The (unit-less) transmembrane
potential assumes values in $[-1,5]$, 
and the resting state of the dynamical system is $(v,n)=(0,0)$.
The function $\cR(n)$ acts as a nonlinear modulator of 
the time-frame between the end of an action potential pulse and the
beginning of the next one (diastolic interval), as well as the duration of the
subsequent action potential pulse. The dispersion map $\cS(n)$ is based on 
experimental restitution properties and it relates
the instantaneous speed of the action potential front-end at a given
spatial point, with the time elapsed since the back-end of a previous
pulse that has passed through the same location. In turn, these
functions are tuned by the parameters $L,M$, respectively.
With the specification \eqref{eq:tauT} we are extending the existing models 
by including an Arrhenius exponential law that modifies the dynamics of the gating variable 
through the function $Q_{10}=\mu^{-(T-T_0)/10}$. This term 
characterises the action of temperature through the mechanism of ionic feedback.
In this expression, $T_0$ represents the reference temperature, i.e. $37^\circ{\rm C}$,
and the law remains valid within a 10-degrees range. 
Furthermore, the so-called Moore term defining the time
constant $\tau_v(T)$ associated to the transmembrane voltage is 
assumed to follow a linear variation with $T$.}

The model from \cite{karma:1994} has been designed specifically 
for cardiac tissue and it has been used in many high-resolution 
2D and 3D electrophysiological studies that match various types 
of experimental data \citep{gizzi:2017}. \cred{A number of more accurate 
physiological cellular models are available from the literature, 
but we restrict to \eqref{eq:reactions} as the complexity in our 
 model resides more in the multi-field coupling framework and in its 
 suitability for large scale electromechanical simulations. 
 Extensions to the two-variable model in \eqref{eq:genKarma} 
 that stay on the phenomenological realm include the three and four-variable 
 systems proposed in \cite{fenton:1998,bueno:2008}, 
 and they provide further experimental validation of the suitability of simplified 
models for the study of a wide class of physiological and pathological scenarios. 
More specific aspects of possible model extensions will be discussed in Section~\ref{sec:concl}.} 
An additional generalisation with respect to \cite{karma:1994} is the self-diffusion 
due to voltage and the account for anisotropy in the diffusion. 
Due to the Piola transformation (forcing a compliance
of the diffusion tensor using the deformation gradients), the
conductivity tensor $\bD(\cdot,\cdot)$ in \eqref{eq:genKarmaV} depends
nonlinearly on the deformation gradient $\bF$, whereas
self-diffusion is here taken as the potential-dependent diffusivity
proposed in \cite{gizzi:2017}, but appropriately modified to incorporate information 
about preferred directions of diffusivity according to the microstructure of the tissue. 
\cred{This model is motivated by diffusion in porous media \citep{vazquez:2006}, 
which has been applied to cardiac tissue in \cite{hurtado:2016},  
justified by the porous nature of the medium \citep{lee:2009} and by the multiscale character 
of diffusion (intercalated discs and gap junctions at the cell level and micro-tubuli at the subcelullar scale, 
\citealp{weinberg:2017}).} More precisely, we set
\begin{equation}\label{def:D}
\bD(v,\bF) = [D_0/2 + D_1 v + D_2 v^2]J\bC^{-1} + D_0/2J\bF^{-1} \cred{\boldsymbol{f}\otimes\boldsymbol{f}} \bF^{-T},
\end{equation}
where \cred{$\boldsymbol{f} = \bF\fo$, and} where the values taken by the parameters $D_i$, $i=0,1,2$ (as well as all
remaining model constants) are displayed in Table~\ref{table:params},
below. Note that here the diffusivity is mainly affected by the fibre-to-fibre connections, 
and 
the presence of $J\bC^{-1}$ suggests a strain-enhanced tissue
conductivity, also referred to as geometric feedback
\citep{colli:2016}. The constants $D_1,D_2$ encode the effect of
linear and quadratic self-diffusion, and they have special importance
at the depolarisation plateau phase, since they modify the speed and
\cred{action potential duration} of the propagating waves. We also note that even for
resting transmembrane potential, the conductivity tensor remains
positive definite.

\begin{figure}[t!]
\begin{center}
\includegraphics[width=0.89\textwidth]{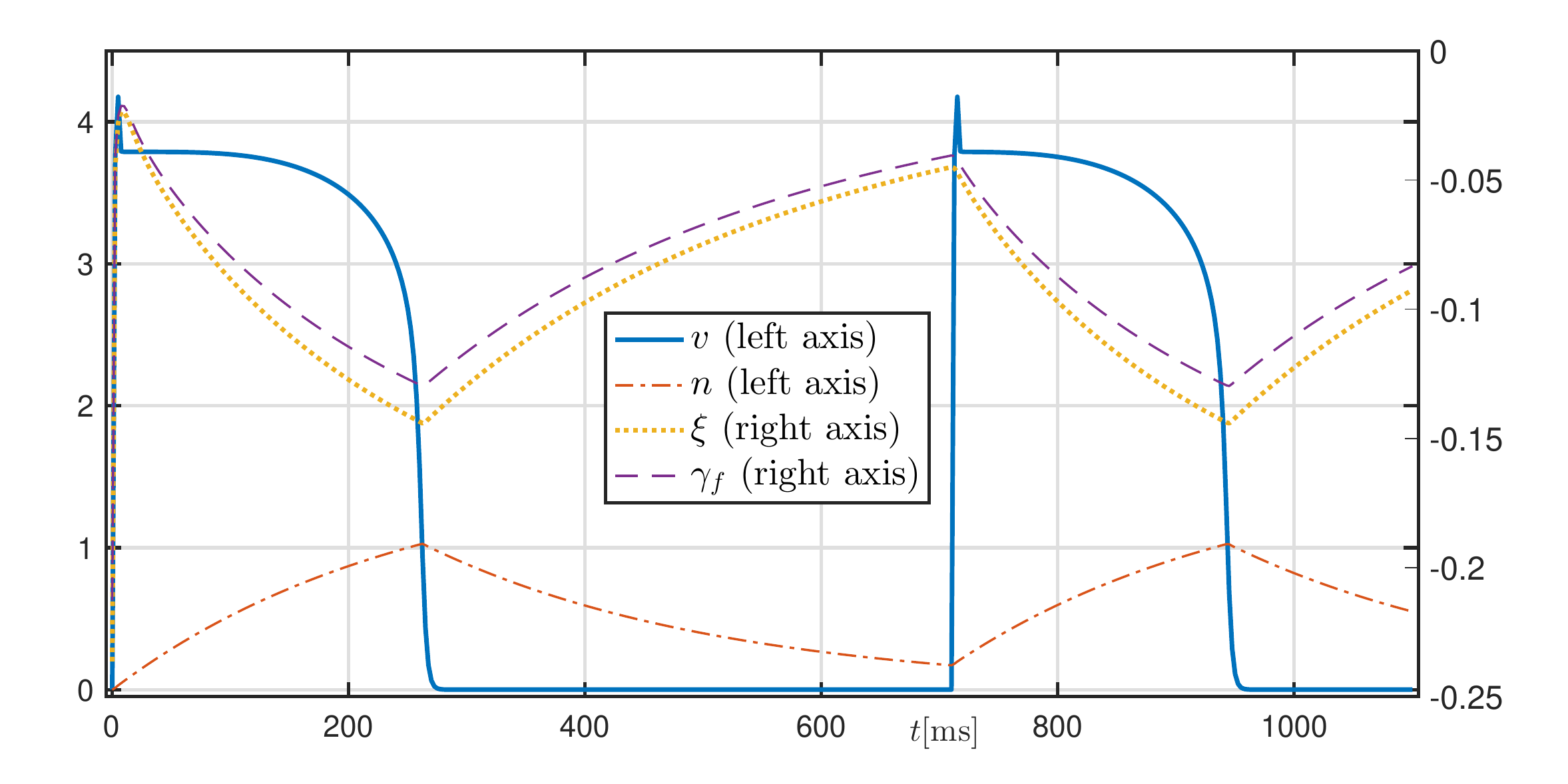}\\
\includegraphics[width=0.325\textwidth]{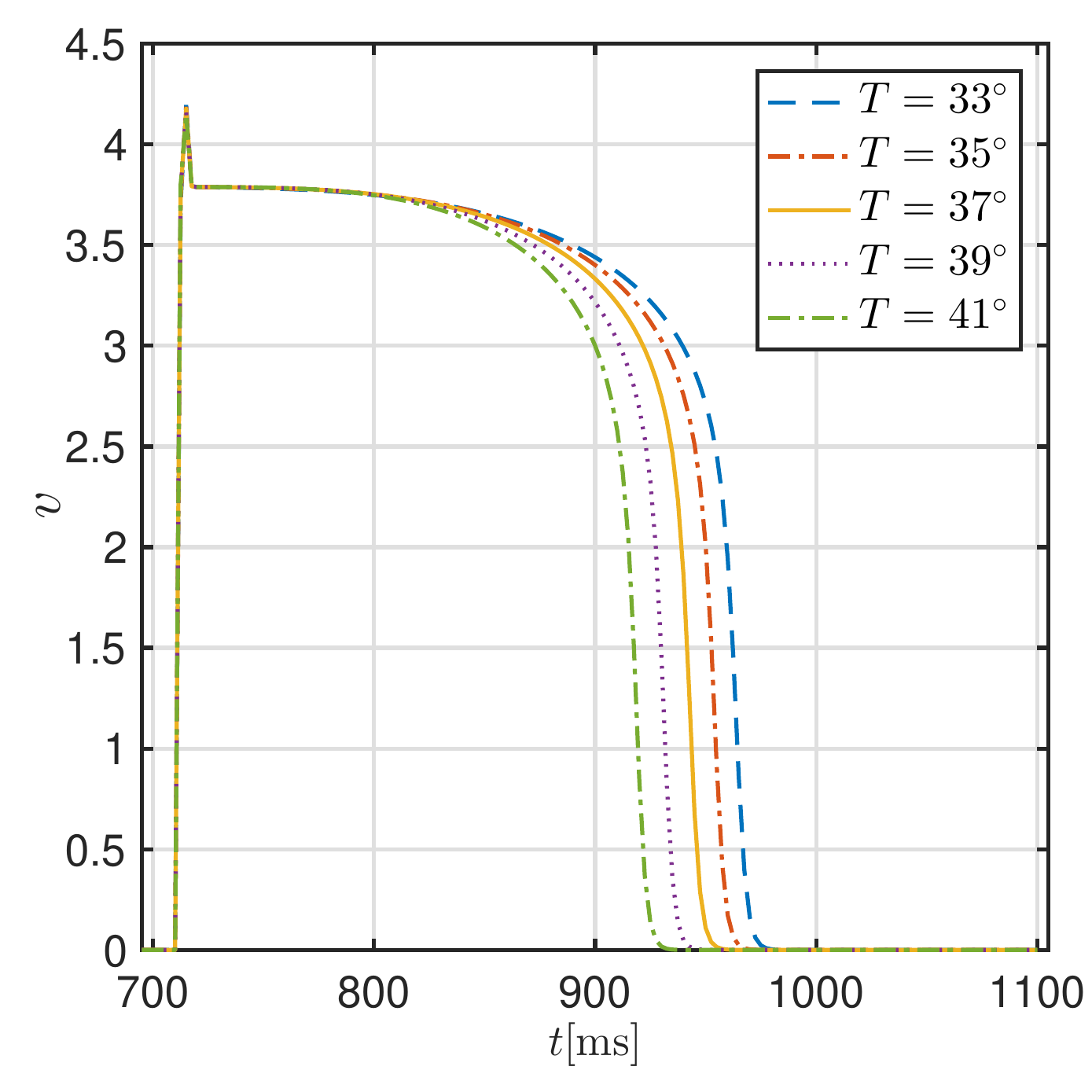}
\includegraphics[width=0.325\textwidth]{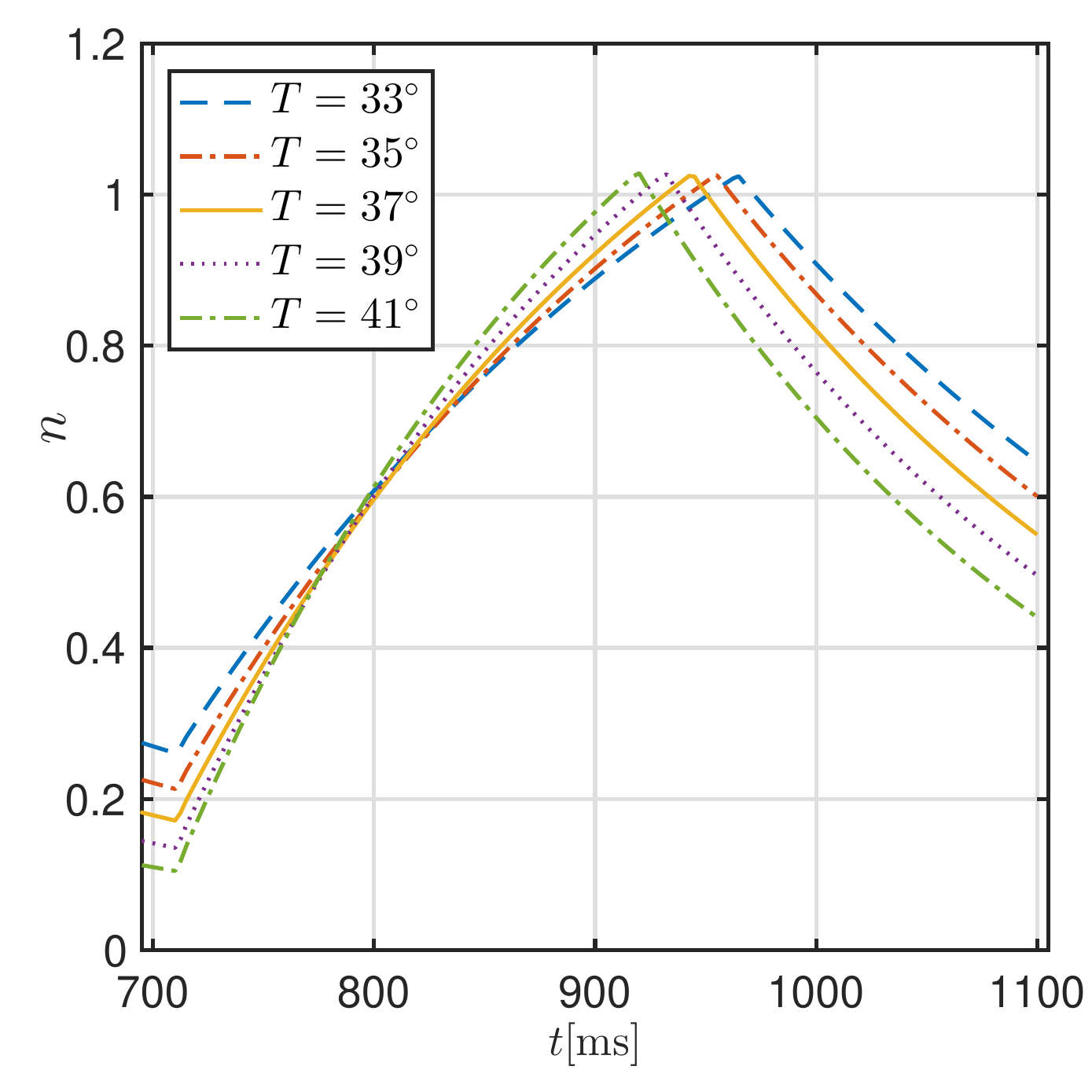}
\includegraphics[width=0.325\textwidth]{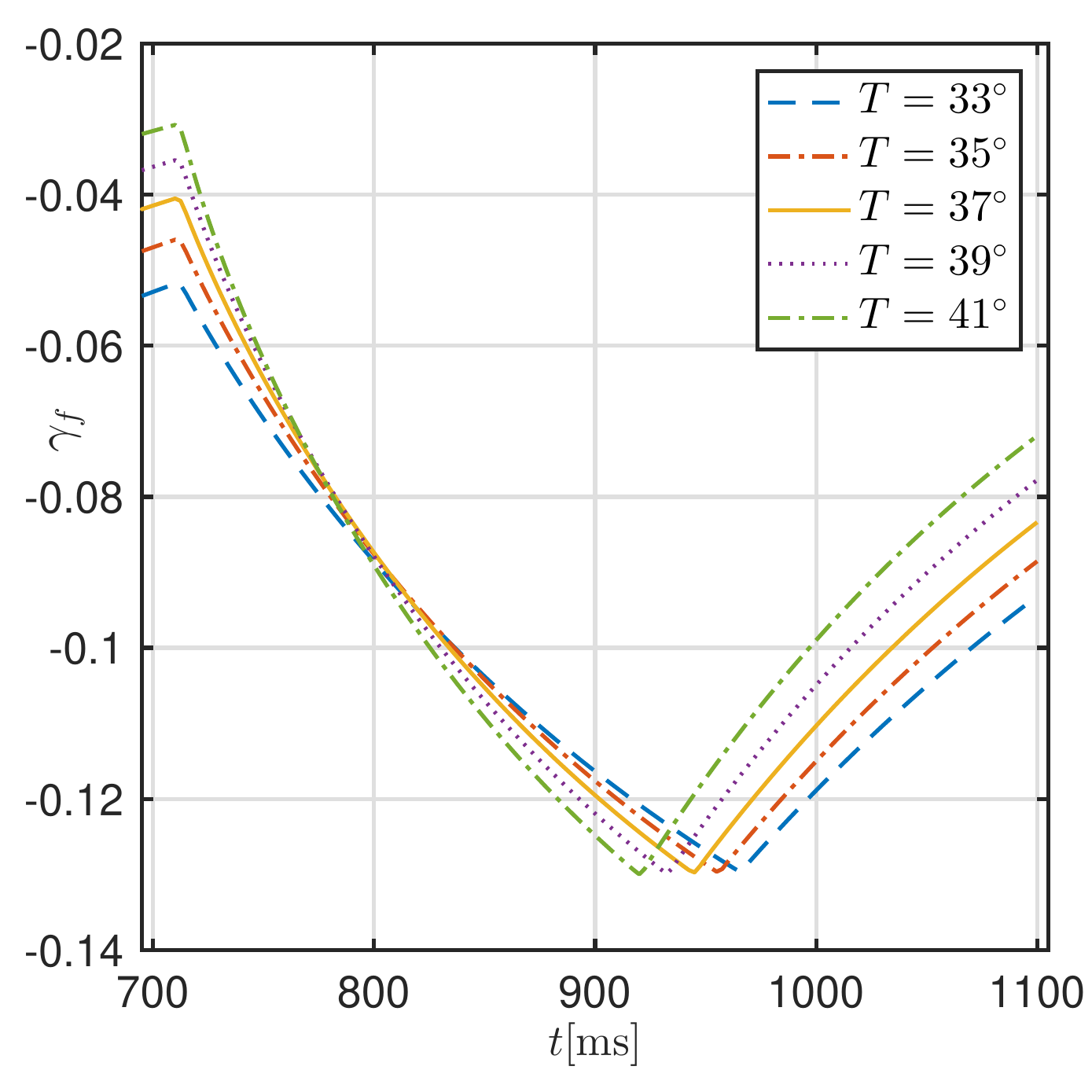}
\end{center}

\vspace{-0.5cm}
\caption{\cred{Top: kinetics of voltage, gating variable (left axis) and cell shortening with 
active strain function (right axis) plotted against time. Bottom: Variations of the dynamics of the coupled thermoelectric model according to 
temperature. The variations of the active strain and myocyte shortening coincide and therefore the 
latter are not shown.}}
\label{fig:kinetics}
\end{figure}

\subsection{Activation mechanisms}\label{sec:activation}
A constitutive equation for the activation functions $\gamma_i$ in 
terms of the microscopic cell shortening $\xi$ is adapted from 
\cite{rossi:2014} as follows 
\begin{equation}\label{eq:gammas}
\gamma_f(\xi) = \cred{\gamma_0} \xi, \quad  \gamma_s(\xi) = (1+\cred{\gamma_0}\xi)^{-1}(1+K_0\cred{\gamma_0}\xi)^{-1}-1, 
\quad \gamma_n(\xi) = K_0\cred{\gamma_0}\xi,
\end{equation}
\cred{where $\gamma_0$ is a positive constant that can 
control the intensity of the activation, and where} the specific relation between the myocyte shortening $\xi$ 
and the dynamics of slow ionic quantities (in the context 
of our phenomenological model, only $n$) is made precise 
using the following law 
\begin{equation}\label{eq:xi}
\frac{d\xi}{dt}  = \dfrac{\ell(\xi,n)}{\tau_\xi (T)} \qquad \text{in } 
\Omega\times(0,t_{\mathrm{final}}],
\end{equation}
where, in analogy to \eqref{eq:reactions}, the dynamics of the 
myocyte shortening are here additionally modulated by 
a temperature-dependent constant 
\begin{equation}\label{eq:h}
\ell(\xi,n):= K_1 \cred{\xi^2} (1+n)^{-1} - K_2 \cred{n} , \qquad 
\tau_\xi(T):=\tau_\xi^0\,
\cred{\tilde{Q}_{10}(T)}.
\end{equation}
\cred{Here $\tilde{Q}_{10}(T)=\tilde{\mu}^{-(T-T_0)/10}$. 
The dynamics of these quantities can be observed in the top 
panel of 
Figure~\ref{fig:kinetics}, for the case of base temperature and 
applying two pacing cycles. Examples of the dynamics of the 
thermo-electric quantities on the second cycle, and for 
varying temperatures are collected in the bottom plots of Figure~\ref{fig:kinetics}.} 
The structure of \eqref{eq:xi} suggests that thermo-electric 
effects could be \cred{similarly} incorporated in other models 
for cellular activation (depending on cross-bridge transitions 
\citep{land:2017}, on calcium - stretch rate couplings in a 
viscoelastic setting \citep{eisner:2017}, or using 
phenomenological descriptions), that is 
through a \cred{phenomenological rescaling with $\tau_\xi$. 
However, a deeper understanding on the precise modification 
of ionic activity in the presence of temperature gradients would be 
a much more difficult task}. 

\subsection{Initial and boundary conditions}
Equations \eqref{eq:momentum}-\eqref{eq:mass} will be supplemented 
either with mixed normal displacement-traction boundary conditions 
\begin{equation}\label{eq:mixedBC}
\bu\cdot \bn = 0 \quad \text{on}\quad \partial\Omega_D\times(0,t_{\mathrm{final}}], \qquad \text{and} \qquad 
\bP \bn = p_N J \bF^{-\tt t}\bn \quad \text{on}\quad \partial\Omega_N\times(0,t_{\mathrm{final}}],
\end{equation}
(where $\partial\Omega_D$,$\partial\Omega_N$ conform a disjoint partition of the boundary, 
the 
traction written in terms 
of the first Piola-Kirchhoff stress tensor is $\bt = \bP\bn$, and 
the term $p_N$ denotes a possibly time-dependent prescribed boundary pressure), 
or alternatively with Robin  conditions on the whole boundary 
\begin{equation}\label{eq:robin}
\bP \bn + \eta J \bF^{-\tt t} \bu = \cero \qquad \text{on}\quad \partial\Omega\times(0,t_{\mathrm{final}}],
\end{equation}
which account for stiff springs 
connecting the cardiac medium with the surrounding soft tissue 
and organs (whose stiffness is encoded in the scalar $\eta$).   On the 
other hand, for the nonlinear diffusion equation 
\eqref{eq:genKarmaV} we prescribe zero-flux boundary conditions
representing insulated tissue 
\begin{equation}\label{eq:zeroflux}
\bD(v,\bF)\, \nabla v \cdot \bn = 0 \qquad \text{on } \partial\Omega\times(0,t_{\mathrm{final}}].
\end{equation}
Finally, the coupled set of equations is closed after defining 
adequate initial data for the transmembrane potential and 
for the internal variables $\xi,n$: 
\begin{equation}\label{eq:initial}
v(\bx,0) = v_0(\bx), \quad n(\bx,0) = n_0(\bx), \quad \xi(\bx,0) = \xi_0(\bx) \qquad \text{on } \Omega\times \{0\}.
 \end{equation}
For the electrical \cred{and activation model we chose resting values for the transmembrane potential, 
the slow recovery, and the myocyte shortening $v_0=n_0 = \xi_0 = 0$}, where initiation of wave propagation will be 
induced with S1-S2-type protocols.

\section{Galerkin finite element method} \label{sec:fe}
\subsection{Mixed-primal formulation in weak form}
The specific structure of the governing
equations (written in terms of the Kirchhoff stress, displacements, solid pressure, electric
potential, activation generation, and ionic variables) suggests to
cast the problem in mixed-primal form, that is, setting the active mechanical 
problem using a three-field formulation, and a primal form for the 
equations driving the electrophysiology. Further details on similar 
\cred{formulations for nearly incompressible hyperelasticity problems} 
can be found in \cite{ruiz:2015,chavan07}. Restricting to the case of 
Robin boundary data for the mechanical problem, 
we proceed 
to test \eqref{eq:momentum}, \eqref{eq:mass}, \eqref{eq:Pi} against adequate 
functions, and doing so also for \eqref{eq:genKarma} yields the problem: For 
$t>0$, find $(\bPi,\bu,p)\in {\mathbb{L}_{\tt sym}^2(\Omega)}\times \mathbf{H}^1(\Omega) 
\times \mathrm{L}^2(\Omega)$ and $(v,n,\xi)\in \mathrm{H}^1(\Omega)^3$ such that 
\begin{subequations}\label{eq:weak-form}
\begin{align}
\int_\Omega [\bPi - \cG(\bu) + pJ\cred{\bI}] : \btau & = 0 & \forall \btau \in {\mathbb{L}_{\tt sym}^2(\Omega)}, \label{eq:weak1}\\
\int_\Omega \bPi : \nabla\bv \bF^{\tt -t} {+} \int_{\partial\Omega} \eta  \bF^{-\tt t} \bu\cdot\bv 
& = \int_\Omega \rho_0 \bb\cdot\bv &  \forall \bv \in \mathbf{H}^1(\Omega),\label{eq:weak2}\\
\int_\Omega [J  - 1 ] q & = 0 &  \forall q\in \mathrm{L}^2(\Omega), \\
\int_\Omega \frac{\partial v}{\partial t} w + \int_\Omega  \bD(v,\bF)\, \nabla v \cdot \nabla w 
  &= \int_\Omega \biggl[\frac{f(v,n)}{\tau_v(T)} + I_{\rm ext}\biggr] w & \forall w \in  \mathrm{H}^1(\Omega),\\
\int_\Omega \biggl(\frac{\partial n}{\partial t} m +  \frac{\partial \xi}{\partial t} \varphi\biggr) & = \int_\Omega 
\biggl(\frac{g(v,n)}{\tau_n(T)} m + \frac{\ell(\xi,n)}{\tau_\xi(T)} \varphi \biggr) & \forall (m,\varphi)\in \mathrm{H}^1(\Omega)^2.
\end{align}\end{subequations}

\subsection{Galerkin discretisation} 
The spatial discretisation
follows a mixed-primal Galerkin approach based on the formulation 
\eqref{eq:weak-form}. \cred{Our} mechanical 
solver constitutes an extension of 
the formulation in \cite{chavan07} to the case of fully incompressible orthotropic materials, whereas  
a somewhat similar method (but using a stabilised form and dedicated to simplicial meshes) has been recently employed 
in \cite{propp:2018} for cardiac viscoelasticity. This family of discretisations has the advantage that 
the incompressibility constraint is enforced in a robust manner. 

Let us denote by $\cT_h$ a regular partition of $\Omega$ 
into hexahedra $K$ of maximum diameter 
$h_K$, and define the meshsize as $h:=\max\{h_K: K\in \cT_h\}$. 
The specific finite element method we chose 
here is based on solving  
 the discrete weak
form of the hyperelasticity equations  using, for the lowest-order case, piecewise constant functions 
to approximate each entry of the symmetric Kirchhoff stress tensor, 
piecewise linear approximation of displacements, and piecewise constant 
approximation of solid pressure. \cred{In turn, all unknowns in the
thermo-electrical model} are discretised with piecewise linear and continuous finite elements. 
More \cred{generally, we can use arbitrary-order} finite dimensional 
spaces $\mathbb{H}_h\!\subset\! {\mathbb{L}_{\tt sym}^2(\Omega)}$, 
$\mathbf{V}_h\subset \bH^1(\Omega)$, $W_h\subset \mathrm{H}^1(\Omega)$, 
$Q_h\subset\mathrm{L}^2(\Omega)$ defined as follows: 
\begin{equation}\label{eq:fespaces}
\begin{split}
\mathbb{H}_h&:=\{ \btau_h\in  {\mathbb{L}_{\tt sym}^2(\Omega) : \tau_h^{ij}\in \mathbb{P}_k(K),  
\ \forall i,j}\in\{1,\ldots,d\},\forall K\in \cT_h\},\\
\mathbf{V}_h&:=\{\bv_h\in \bH^1(\Omega):\bv_h|_K\in\mathbb{P}_{k+1}\cred{(K)^3},\forall  K\in \cT_h\},\\
Q_h&:=  \{ q_h\in \mathrm{L}^2(\Omega) : q_h|_K\in \mathbb{P}_{k}(K),\forall  K\in \cT_h\},\\
W_h&:=  \{ w_h\in \mathrm{H}^1(\Omega) : w_h|_K\in \mathbb{P}_{k+1}(K),\forall  K\in \cT_h\},
\end{split}\end{equation}
where $\mathbb{P}_r(K)$ denotes the space of polynomial functions of degree $s\leq r$ 
defined {on the hexahedron $K$}. 
Assuming zero body loads, and applying a backward Euler time integration we end up with the following fully-discrete 
nonlinear electromechanical problem, starting from the discrete initial data 
$v^0_h,n^{0}_h,\xi^0_h$. 
For each $j=0,1,\ldots$: find $(\bPi_h^{j+1},\bu_h^{j+1},p_h^{j+1})$ and $(v_h^{j+1},n_h^{j+1},\xi_h^{j+1})$ 
such that 
\begin{subequations}\label{eq:galerkin}
{\small\begin{align}
\int_\Omega [\bPi_h^{j+1} - \cG(\bu_h^{j+1}) + p_h^{j+1}J(\bu_h^{j+1})\cred{\bI}] : \btau_h & = 0 & \forall \btau_h \in \mathbb{H}_h,\label{galerkin-Pi}\\
\int_\Omega \bPi_h^{j+1} : \nabla\bv_h \bF^{\tt -t}(\bu_h^{j+1}) {+} \int_{\partial\Omega} \eta \bF^{\tt -t}(\bu_h^{j+1}) \bu_h^{j+1}\cdot\bv_h 
& = 0 &  \forall \bv_h \in \mathbf{V}_h,\label{galerkin-u}\\
\int_\Omega [J(\bu_h^{j+1})  - 1 ] q_h & = 0 &  \forall q_h\in Q_h, \label{galerkin-p}\\
\int_\Omega \frac{v_h^{j+1}-v_h^{j}}{\Delta t} w_h + \int_\Omega  \bD(v_h^{j+1},\bF(\bu_h^{j+1}))\, \nabla v_h^{j+1} \cdot \nabla w_h 
  - \int_\Omega \biggl[\frac{f(v_h^{\cred{j}},n_h^{\cred{j}})}{\tau_v(T)} + I_{\rm ext}\biggr] w_h &= 0 & \forall w_h \in  W_h,\label{galerkin-v}\\
\int_\Omega \frac{n_h^{j+1}-n_h^{j}}{\Delta t} m_h  - \int_\Omega 
\frac{g(v_h^{\cred{j}},n_h^{\cred{j}})}{\tau_n(T)} m_h  & = 0 & \forall m_h \in  W_h,\label{galerkin-n}\\
\int_\Omega   \frac{\xi_h^{j+1}-\xi_h^{j}}{\Delta t} \varphi_h  - \int_\Omega 
\frac{\ell(\xi_h^{\cred{j}},n_h^{\cred{j}})}{\tau_\xi(T)} \varphi_h  & = 0 & \forall \varphi_h\in  W_h.\label{galerkin-xi}
\end{align}}\end{subequations}
\cred{Due to the intrinsic interpolation properties of the finite-dimensional spaces specified in \eqref{eq:fespaces}, we expect to observe $\mathcal{O}(h^{k+1})$ convergence for Kirchhoff stress and pressure in the tensor and scalar $L^2-$norms, as well as $\mathcal{O}(h^{k+1})$ convergence for the remaining fields in the $\mathbf{H}^1-$norm (which reduces to first-order convergence for the lowest-order finite element family, $k=0$).} 

 \cred{Alternatively to the method above, if we do not apply 
integration by parts in  
 \eqref{eq:weak2}, one can redefine a method that seeks for  
$\mathbb{H}(\mathbf{div};\Omega)$-conforming approximations for the 
Kirchhoff stress and $\mathbf{L}^2(\Omega)$ - conforming approximations 
of displacements. That is, for instance using Raviart-Thomas 
elements of first order to approximate rows of the Kirchhoff stress tensor, 
and piecewise constant approximation of displacements \citep{gatica:book}, appropriately 
modified for the case of hexahedral meshes.}

\begin{table}[t]
\setlength{\tabcolsep}{2pt}
\centering{}
\cred{\small\begin{tabular}{rlclrlclrlclrlc}
\hline\noalign{\smallskip} 
\hline\noalign{\smallskip} 
\multicolumn{15}{c}{Thermo-electric model parameters}\\
\hline\noalign{\smallskip}
$v_\star=$   & 3 & [--]&&\quad $v_n=$ & 1 & [--] & & \quad $v^*=$   & 1.5415 &[--] &&\quad $ \mu=$ & 1.5 & [--]\\
${\tau^0_v}=$& 2.5 & [ms] &&\quad $\beta=$ & 0.008 & [--] & & \quad ${\tau^0_n}=$& 250 & [ms] &&\quad $D_0=$ & 0.85 & [cm$^2$/s]\\
$D_1=$ & 0.09 & [cm$^2$/s] &&\quad $D_2=$ & 0.01 & [cm$^2$/s] && \quad $L=$    & 0.9 & [--] & & \quad $M=$     & 9 &[--] \\
 $T_0=$ & 37 & [$^\circ$C] &&\quad $ \tilde{\mu}=$ & 3.9 & [--]\\
\hline\noalign{\smallskip}
\multicolumn{15}{c}{Mechano-chemical model parameters}\\
\hline\noalign{\smallskip}
$a=$    & 0.333 & [kPa] && \quad $a_{f}=$ & 18.535 & [kPa] & & \quad $a_{s}=$ & 2.564 & [kPa] && \quad $a_{fs}=$ & 0.417 & [--] \\
$b=$    & 9.242& [--]        && \quad $b_{f}=$ & 15.972 & [--] & & \quad $b_{s}=$ & 10.446& [--] && \quad $b_{fs}=$ & 11.602 & [--] \\
$K_0=$   & 5 &[--] &&\quad $K_1=$ & 3.5 & [--]&&\quad $K_2=$ & 0.035 & [--] & & \quad $ \eta \in$     & \{0.05,0.9\} &[kPa]  \\
${\tau^0_\xi=}$& 0.5 & [ms] & & \quad $\gamma_0=$ & 0.9 & [--] \\
\hline\noalign{\smallskip}
\hline\noalign{\smallskip}
\end{tabular}
}
\caption{\label{table:params}
Coefficients for the electromechanical model \eqref{eq:mech}, \eqref{eq:genKarma}, \eqref{eq:xi}, with values 
 taken from \cite{cherubini:2017,holzapfel:2009,rossi:2014}.}
\end{table}

\begin{algorithm}[t!]
\algsetup{indent=2em}
\caption{-- Overall coupled electromechanics}
\label{alg:EM}
\cred{\small\begin{algorithmic}[1]
\FOR {a given computation start with an offline phase \AND}
	\STATE \textbf{Set} geometry, size and orientation, and \textbf{assign} boundary labels to the epicardium $\partial\Omega_{\text{epi}}$, endocardium  $\partial	\Omega_{\text{endo}}$, and basal cut  $\partial\Omega_{\text{base}}$
	\STATE \textbf{Define} a global meshsize and \textbf{construct} hexahedral meshes (surface and volumetric)
	\STATE \textbf{Generate} orthotropy of the medium through the rule-based algorithm in mixed form 
 	\STATE \textbf{Set} maximal and minimal angles for rotational anisotropy  $\theta_{\text{epi}}$ and $\theta_{\text{endo}}$ 
         \STATE \textbf{Set} the ventricular centreline vector $\ko$ 
          \STATE \textbf{Define} mixed finite-dimensional spaces for the approximation of a potential $\phi$ and an auxiliary sheetlet field $\bzeta$
          \STATE \textbf{Apply} boundary conditions on the finite element space for sheetlets \AND solve the discrete counterpart of \eqref{eq:problem-fibres}
          \STATE \textbf{Obtain} sheetlet directions from $\so = \bzeta_h / \|\bzeta_h\|$
           \STATE \textbf{Project} the centreline as follows $\widehat{\ko} = \ko - (\ko \cdot \so) \so$   
           \STATE \textbf{Compute} flat fibres field $\widehat{\fo} = \so \times \widehat{\ko}/\|\widehat{\ko}\|$
           \STATE \textbf{Apply} a rotation of flat fibres incorporating intramural angle variation
\ENDFOR          
\STATE \textbf{Set} timestep $\Delta t$, initial and final times $t=t_0,t_{\text{final}}$;
\STATE \textbf{Define} mixed finite-dimensional spaces in \eqref{eq:fespaces}
\STATE \textbf{Define} constant and solution-dependent model coefficients
\STATE \textbf{Apply} boundary conditions and set initial solutions from expressions or data
\STATE \textbf{Construct} functional forms appearing in the Galerkin discretisation \eqref{eq:galerkin} 
	\WHILE{ $t < t_{\text{final}}$ }
		\STATE \textbf{Construct} the nonlinear algebraic system associated with \eqref{galerkin-v}-\eqref{galerkin-xi}, 
		taking the reaction terms explicitly 
		\STATE \textbf{Construct} the linear system arising  from the Jacobian of the nonlinear problem
		\FOR {$k=1$ until convergence}
			\STATE \textbf{Assemble} \AND \textbf{solve} the matrix system associated with the Jacobian 
			\STATE \textbf{Update} Newton approximation \AND reinitialise
		\ENDFOR
		\STATE \textbf{Update} thermo-electric solutions  $v_h^{j} \gets v_h^{j+1}$,  $n_h^{j} \gets n_h^{j+1}$,  $\xi_h^{j} \gets \xi_h^{j+1}$
		\AND time-dependent coefficients (e.g. boundary pressure $p_N(t)$)
		\STATE \textbf{Compute} orthotropic activation quantities from \eqref{eq:gammas}
		\STATE \textbf{Construct} the nonlinear algebraic system associated with  \eqref{galerkin-Pi}-\eqref{galerkin-p}
		\STATE \textbf{Construct} the linear system arising  from the Jacobian of the nonlinear problem
		\FOR {$k=1$ until convergence}
		              \STATE \textbf{Assemble} \AND \textbf{solve} the tangent linear system for increments
		              \STATE \textbf{Update} Newton approximation \AND reinitialise
		\ENDFOR
		\STATE \textbf{Update} time: $t\gets t + \Delta t$, $j \gets j + 1$
		\STATE \textbf{Output} solutions for visualisation and data analysis
	\ENDWHILE    
    \end{algorithmic}}
\end{algorithm}

\subsection{Implementation details}
The coupling between activated mechanics and the electrophysiology
solvers is not done monolithically, but rather realised using a segregated fixed-point scheme. The
nonlinear mechanics are solved using an embedded Newton-Raphson method
and an operator splitting algorithm separates an implicit diffusion
solution (where another Newton iteration handles the nonlinear
self-diffusion) from an explicit reaction step for the kinetic
equations, turning the overall solver into a semi-implicit
method. Updating and storing of the internal variables $\xi$ and $n$
is done locally at the quadrature points.  The routines are
implemented using the finite element library FEniCS \citep{fenics},
and in all cases the solution of linear systems is carried out with
the BiCGStab method preconditioned with an algebraic multigrid solver 
(both provided by the PETSc library), 
\cred{and using a relative tolerance of 1e-4 for the unpreconditioned $\ell^2$-norm 
of the residual.}
The domains to be studied consist of 3D slabs, \cred{ring-shaped}, 
and ellipsoidal geometries with varying thickness and basal cuts, 
discretised into hexahedral meshes of maximum meshsize ${h} =
0.01$\,cm.  The time discretisation uses a fixed timestep $\Delta t$ 
(dictated by the dynamics of the cell ionic model rather 
than by a CFL condition, as the diffusion is discretised implicitly),
and we observe that the hyperelasticity equations have a different
inherent timescale, so we update their solution every five steps taken
by the electrophysiology solver.  Since in \eqref{eq:xi} the evolution
of myocyte shortening does not depend locally on the macroscopic
stretch, the activation system can be conveniently solved together
with the ionic model. A tolerance of 1e-7 on the $\ell^\infty$-norm of
the residual is employed to terminate the Newton iterates 
\cred{for the nonlinear diffusion and for the nonlinear hyperelasticity sub-problems. 
A summary of the overall process, including all steps from 
mesh generation to solution visualisation, is outlined in Algorithm~\ref{alg:EM}.}

\section{Numerical results}\label{sec:results}

\begin{figure}[t!]
\begin{center}
\includegraphics[height=0.42\textwidth]{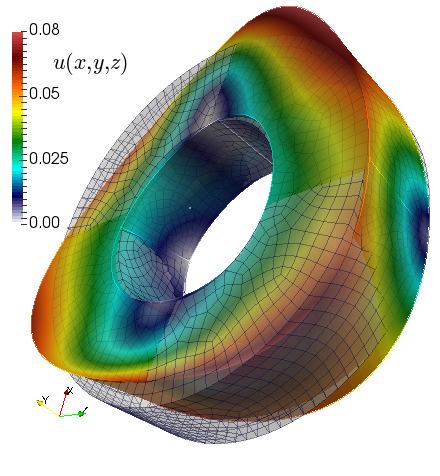}
\includegraphics[height=0.42\textwidth]{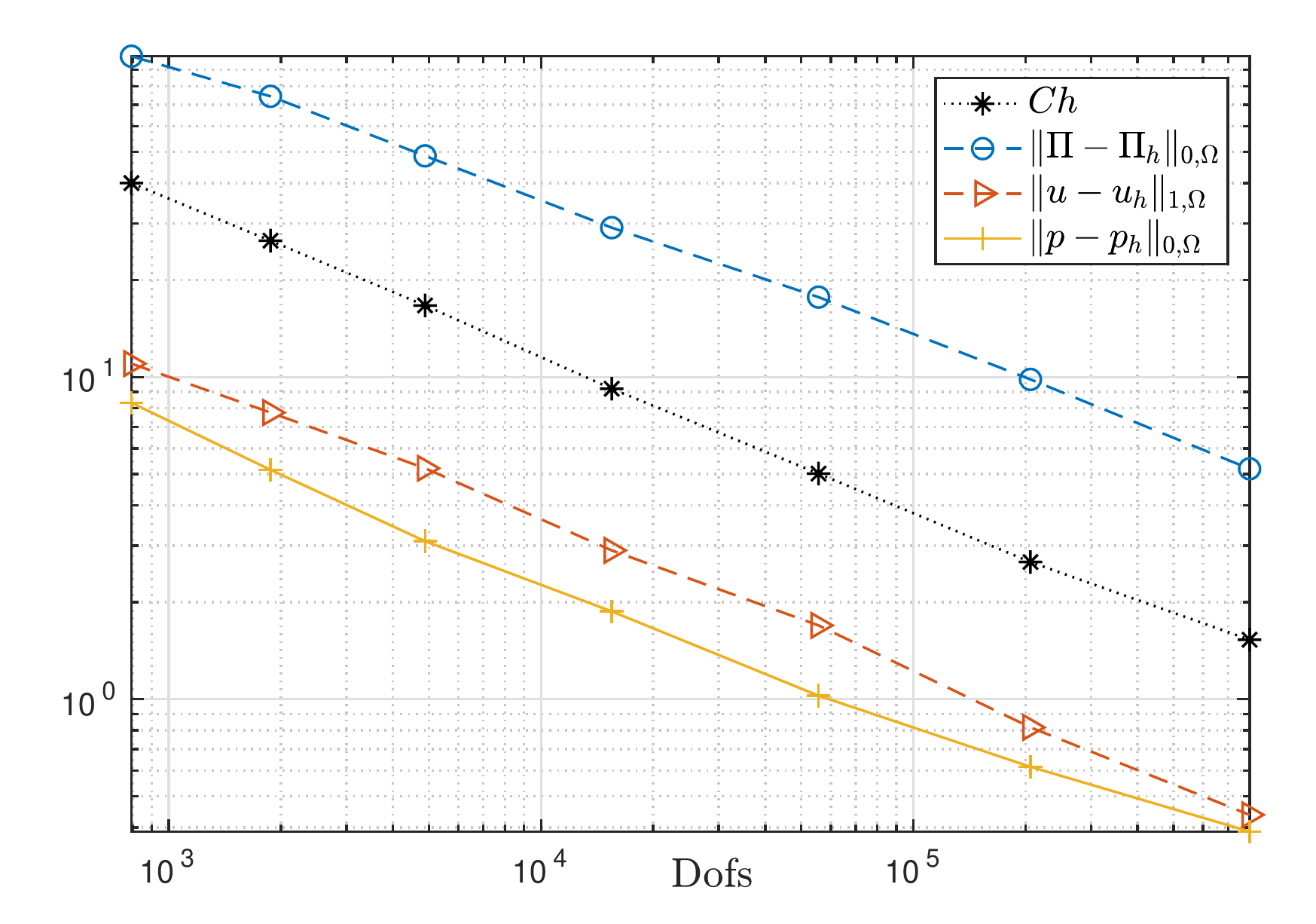}
\end{center}

\vspace{-0.5cm}
\caption{\cred{Example of approximate displacement field on the deformed domain and a coarse hexahedral mesh in the 
undeformed configuration (left); and error history for an accuracy test of the mixed formulation for hyperelasticity (right) 
generated using a lowest-order discretisation.}}
\label{fig:ring}
\end{figure}

\cred{Before carrying out model validation and performing simulations with the fully coupled model 
described in Section~\ref{sec:model}, we conduct a mesh convergence test to assess the accuracy 
of the mixed finite element scheme proposed for the three-field hyperelasticity subproblem \eqref{eq:mech}-\eqref{eq:Pi}, 
in the case when the material is completely passive. The following test has been employed (for isotropic, Mooney-Rivlin materials) as a benchmark for different finite element solvers \citep{chamberland10}. In the 3D domain defined by a ring-shaped region of
 width 0.25\,cm, internal diameter of 0.5\,cm and external diameter of 1\,cm, we define closed-form 
 manufactured solutions as  
$$ p(x,y,z) = x^4 - y^4 - z^4, \quad \bu (x,y,z) =  \biggl(x^4 + \frac{2}{5}yz,\ y^4+ \frac{2}{5}xz,\ \frac{1}{10}z^4-\frac{2}{5}xyz\biggr)^{\tt t},$$
and construct an exact form of the Kirchhoff stress, as well as body loads and eventually traction terms using these 
smooth functions. Sheetlets are radially defined, whereas fibres are clockwise oriented with respect to the $y$ axis, and 
the hyperelasticity parameters are set according to the second part of Table~\ref{table:params}. 
Boundary conditions were considered of mixed type as in \eqref{eq:mixedBC}, 
but setting appropriate non-homogeneous terms. The traction boundary $\partial\Omega_N$ 
corresponds to the  top and bottom faces of the ring (parallel to the $xz$ axis where the normal vector is 
$\bn = (0,\pm 1,0)^{\tt t}$), whereas the 
normal displacement boundary $\partial\Omega_D$ is conformed by the internal and external curved surfaces.  
We compute errors between the exact solutions and the approximate fields generated by the lowest-order 
scheme on a sequence of unstructured hexahedral meshes of different resolutions. These (absolute) errors are measured 
in the tensor and scalar $L^2-$norms for the Kirchhoff stress and pressure, respectively; and in the $\mathbf{H}^1-$norm for the displacements. We plot the results versus the number of degrees of freedom in Figure~\ref{fig:ring}(right), where we can observe an optimal convergence (first-order in this case), as anticipated 
in Section~\ref{sec:fe}. The number of Newton iterates required to reach convergence was in average 4.}

\subsection{Conduction velocity assessment}
\begin{table}[t]
\setlength{\tabcolsep}{4pt}
\bigskip{}\centering{}
\cred{\small\begin{tabular}{cccc}
\hline\noalign{\smallskip} 
\hline\noalign{\smallskip}
Temp.  & $h = 0.025$\,{cm},$\Delta t = 0.03$\,{ms} & $h = 0.0125$\,{cm},$\Delta t = 0.0075$\,{ms} & 
    $h = 0.006$\,{cm},$\Delta t = 0.00125$\,{ms} \\
\hline\noalign{\smallskip}
$T=33^\circ$C & 0.356 & 0.377 & 0.422 \\
$T=35^\circ$C & 0.428 & 0.435 & 0.441 \\
$T=37^\circ$C & 0.439 & 0.447 & 0.453 \\
$T=39^\circ$C & 0.442 & 0.450 & 0.448 \\   
$T=41^\circ$C & 0.443 & 0.451 & 0.451 \\   
\hline\noalign{\smallskip}
\hline\noalign{\smallskip}
\end{tabular}}
\caption{\label{table:conduction}
Computed conduction velocities $[{\rm m/s}]$  according to different temperature values 
and spatio-temporal refinement.}
\end{table}

\begin{figure}[t!]
\begin{center}
\includegraphics[width=0.325\textwidth]{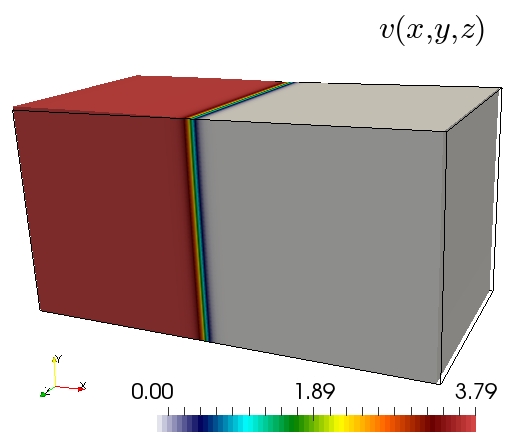}
\includegraphics[width=0.325\textwidth]{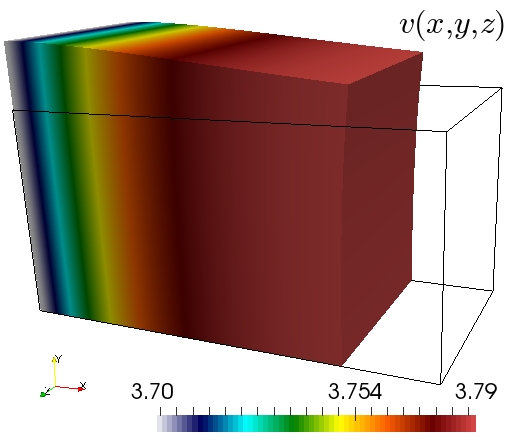}
\includegraphics[width=0.325\textwidth]{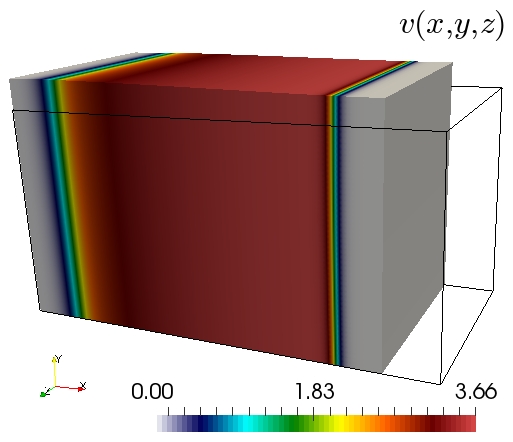}\\
\includegraphics[width=0.325\textwidth]{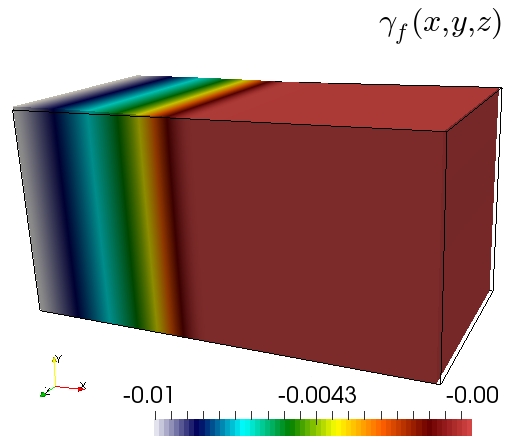}
\includegraphics[width=0.325\textwidth]{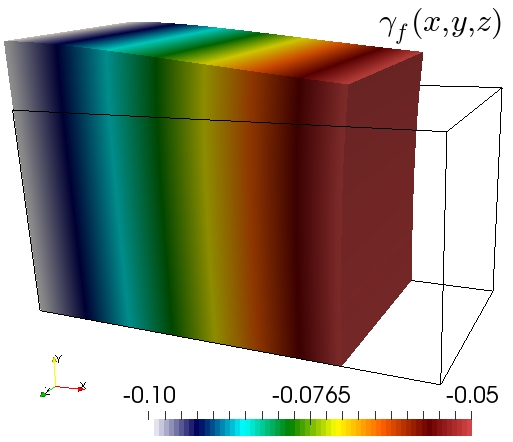}
\includegraphics[width=0.325\textwidth]{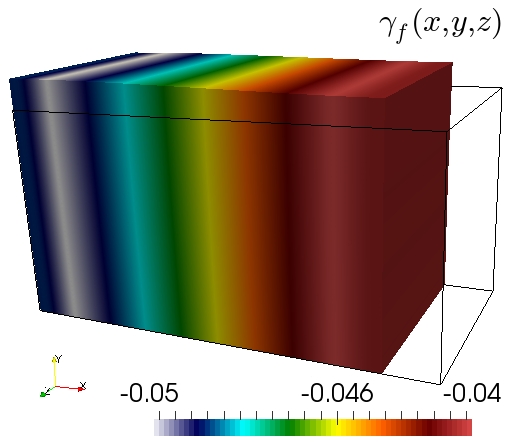}\\
\includegraphics[width=0.325\textwidth]{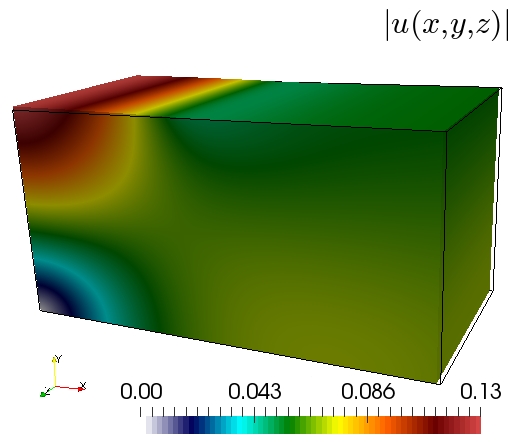}
\includegraphics[width=0.325\textwidth]{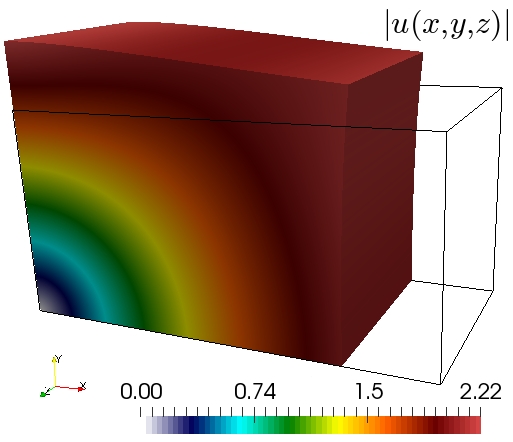}
\includegraphics[width=0.325\textwidth]{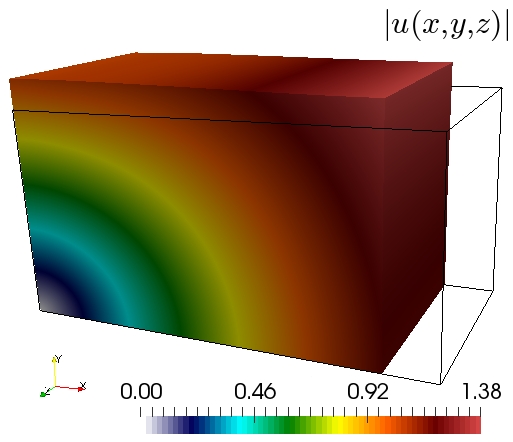}\\
\includegraphics[width=0.325\textwidth]{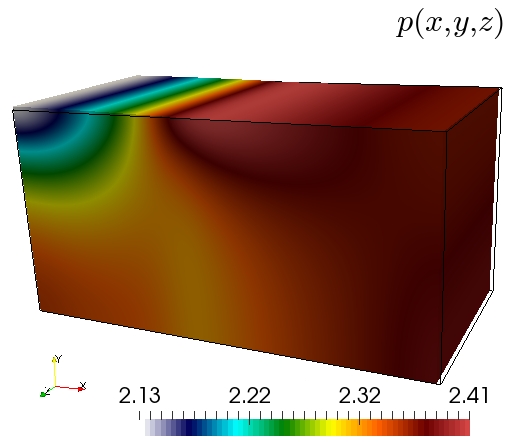}
\includegraphics[width=0.325\textwidth]{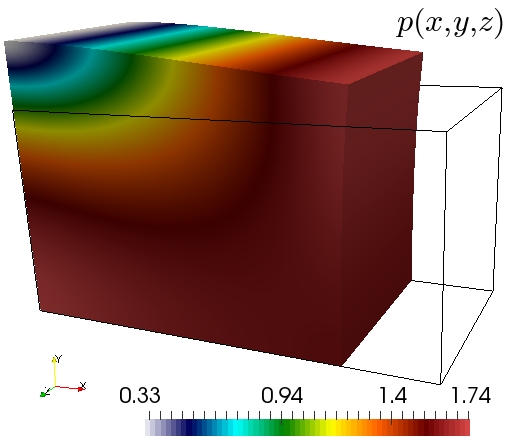}
\includegraphics[width=0.325\textwidth]{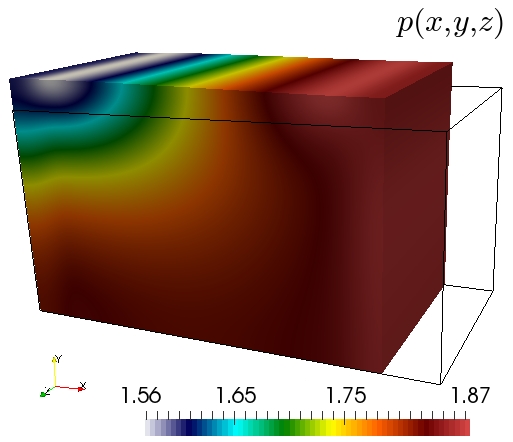}
\end{center}

\vspace{-0.5cm}
\caption{Samples  of the approximate solutions 
(potential, activation, displacement magnitude, and solid pressure) 
shown on the deformed domain at \cred{$t=50,150,450$\,ms (left, middle, and 
right panels)}. For this test we have used $T=39^\circ$C. }
\label{fig:ex01}
\end{figure}

\begin{figure}[t!]
\begin{center}
\includegraphics[width=0.495\textwidth]{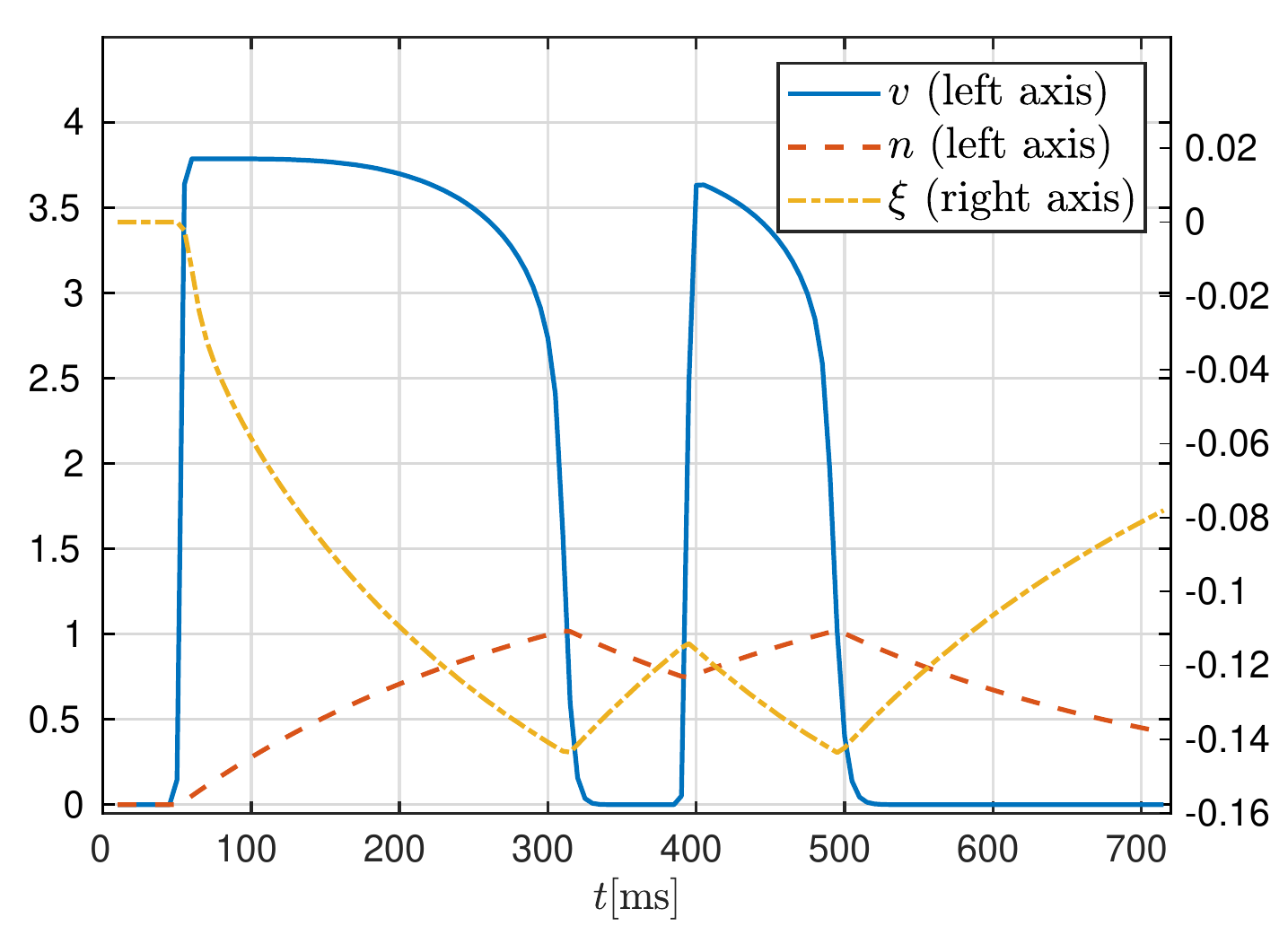}
\includegraphics[width=0.495\textwidth]{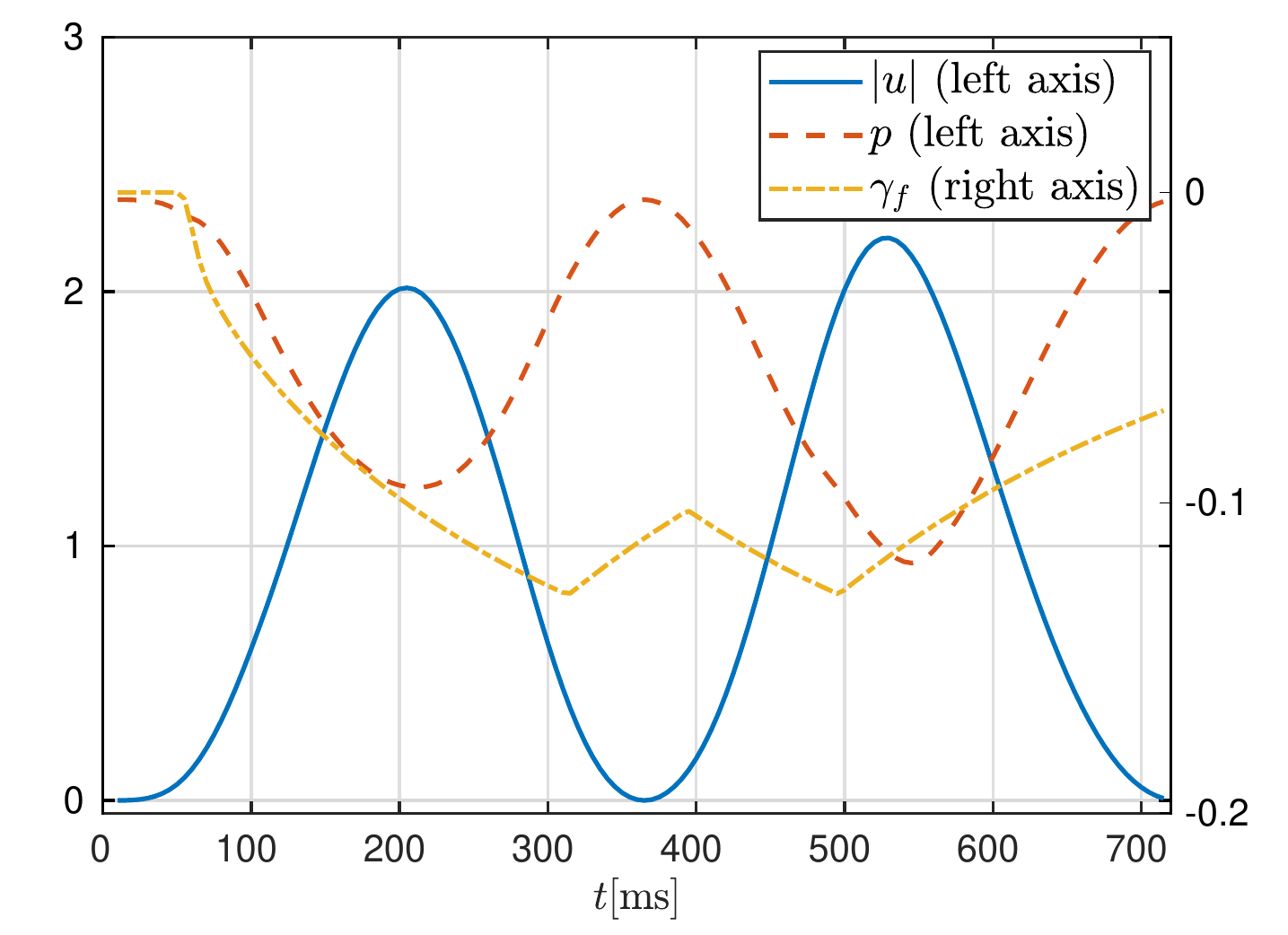}
\end{center}

\vspace{-0.5cm}
\caption{\cred{Evolution of main variables measured on the point (5,2.5,2.5) and up to $t=720$\,ms, 
computed at temperature $T=39^\circ$C.} }
\label{fig:ex01oneP}
\end{figure}

We next consider the electromechanical model 
 \eqref{eq:mech}, \eqref{eq:genKarma}, \eqref{eq:xi} defined on the 3D slab 
\cred{$\Omega =(0,10)\times(0,5)\times(0,5)$\,cm$^3$}.  The boundary conditions correspond to 
\eqref{eq:mixedBC} and \eqref{eq:zeroflux}. The bottom ($z=0$), back ($y=0$), and left ($x=0$) 
sides of the block will constitute $\partial\Omega_D$ where we impose zero normal displacements, 
and on the remainder of the boundary $\partial\Omega_N=\partial\Omega\setminus\partial\Omega_D$ 
we prescribe zero traction.  \cred{We consider only constant 
fibre and sheet directions $\fo = (1,0,0)^{\tt t}$, $\so = (0,1,0)^{\tt t}$}, and 
a stimulus of amplitude \cred{2} and duration 2\,ms is applied on the left 
wall at time $t=1\,$ms, which initiates a planar wave propagation. At \cred{a temperature of} $T=37^\circ$C, the 
thermo-electric effects are turned off (both $Q_{10}$ and Moore terms equal 1), and the reported 
maximum conduction velocity of \cred{45.1\,cm$/$s} can be computed using $\tilde{D}=1.1\,$cm$^2/$s (that 
is, setting $D_1=D_2=0$). Then, 
variations 
of temperature and of the constants that characterise the nonlinear diffusion lead to slight 
modifications on the conduction velocity. Here this value is computed using the approximate 
potential and activation times measured between the points 
\cred{$(x,y,z)=(4.663,2.5,2.5)$ and $(x,y,z)=(5.337,2.5,2.5)$}, that is a spatial variation 
in the $x-$direction of $\delta x=0.674$\,cm, 
and employing a threshold of amplitude 1. We also vary the mesh resolution and observe that the coarsest spatio-temporal discretisation 
 that maintains conduction velocities in physiological ranges requires a meshsize 
of \cred{$h=0.025$\,cm} and a timestep of $\Delta t = 0.03\,$ms. Our results are summarised in Table~\ref{table:conduction}. 
We can note that for the lowest temperatures, the changes in the mesh resolution entail substantial modifications in the conduction velocity, whereas for higher temperatures the effect seems to be milder and even coarse meshes give physiological results. 
After computing each conduction velocity value, we have commenced another pacing 
cycle \cred{(with an S2 applied at $t=330$\,ms), and run the simulation until $t=720$\,ms. 
Snapshots of the potential, activation, displacement magnitude, and solid pressure are depicted in Figure~\ref{fig:ex01}, where we 
can observe (in particular for $t=150$\,ms) a marked deformation in the sheetlet direction complying with the 
shortening in the fibre direction.   In Figure~\ref{fig:ex01oneP} we plot the history 
of the main thermo-electric and kinematic variables on the midpoint of the 
line where conduction velocities are computed.}  
We remark that the different thermal states, in addition to modifying the conduction
velocity, also affect the shape and duration of the action potential wave.  
In agreement with the constitutive modelling, the amount of contraction is not linked to the velocity of propagation but rather to the \cred{duration} of the action potential. More precisely, since the active-strain contraction is linked to the amount of tissue undergoing a certain level of voltage, it turns out that at lower temperature the action potential wave is larger (experimental evidence for this phenomenon can be found in \citealp{fenton:2013}), and therefore the amount of tissue undergoing contraction is larger. 

\subsection{Scroll wave dynamics and localised temperature gradients}
\begin{figure}[t]
\begin{center}
\includegraphics[width=0.495\textwidth]{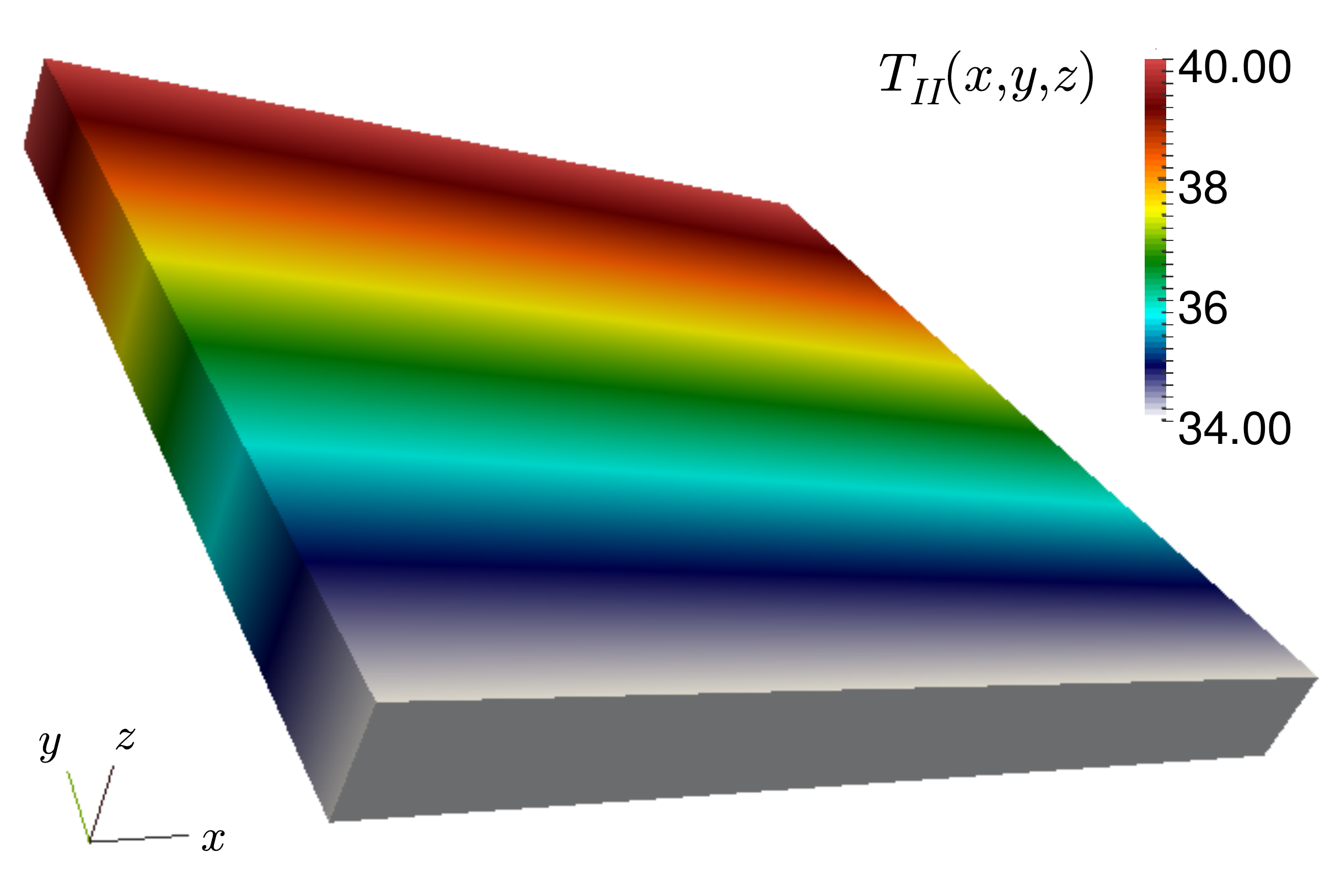}
\includegraphics[width=0.495\textwidth]{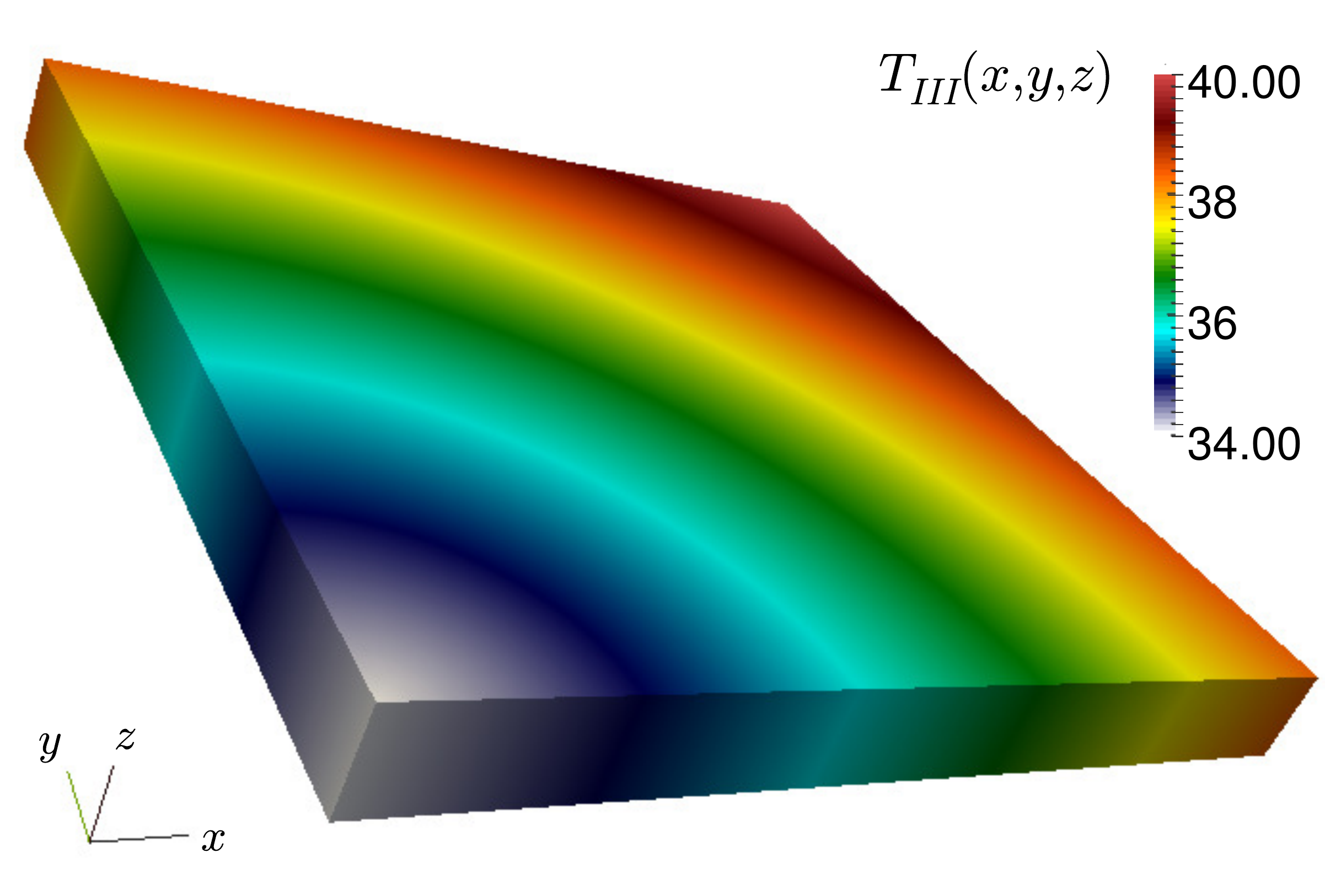}
\end{center}

\vspace{-1.5cm}
\begin{align*}
r = (x^2+y^2)^{1/2},\,\, R = 6.72\sqrt{2},\\
T_{II}(x,y,z) =\frac{1}{6.72}(40^\circ\text{C} \, y+34^\circ\text{C} \, (6.72-y)),\quad 
T_{III}(x,y,z)  =\frac{1}{R}(40^\circ\text{C} \,r+34^\circ\text{C} \, (R-r)).
\end{align*}

\vspace{-0.5cm}
\caption{Temperature distributions in the undeformed configuration, where the  
colour code is in $^\circ$C.}
\label{fig:ex02temp}
\end{figure}

We now perform a series of tests aimed at analysing the 
differences in wave propagation patterns produced with different 
temperature conditions such as those encountered in transmural gradients 
induced by fever, cold/hot water, and/or localisation of other thermal sources such as 
ablation devices. First on the case of the base temperature 
$T=37^\circ$C, secondly in the case where the domain is 
subject to a temperature gradient in the direction of the sheetlets 
$\so=(0,1,0)^{\tt t}$, and third when the temperature has a radial gradient in the $xy$ 
plane (\cred{see Figure~\ref{fig:ex02temp}}). 
  
\begin{figure}[t]
\begin{center}
\includegraphics[width=0.495\textwidth]{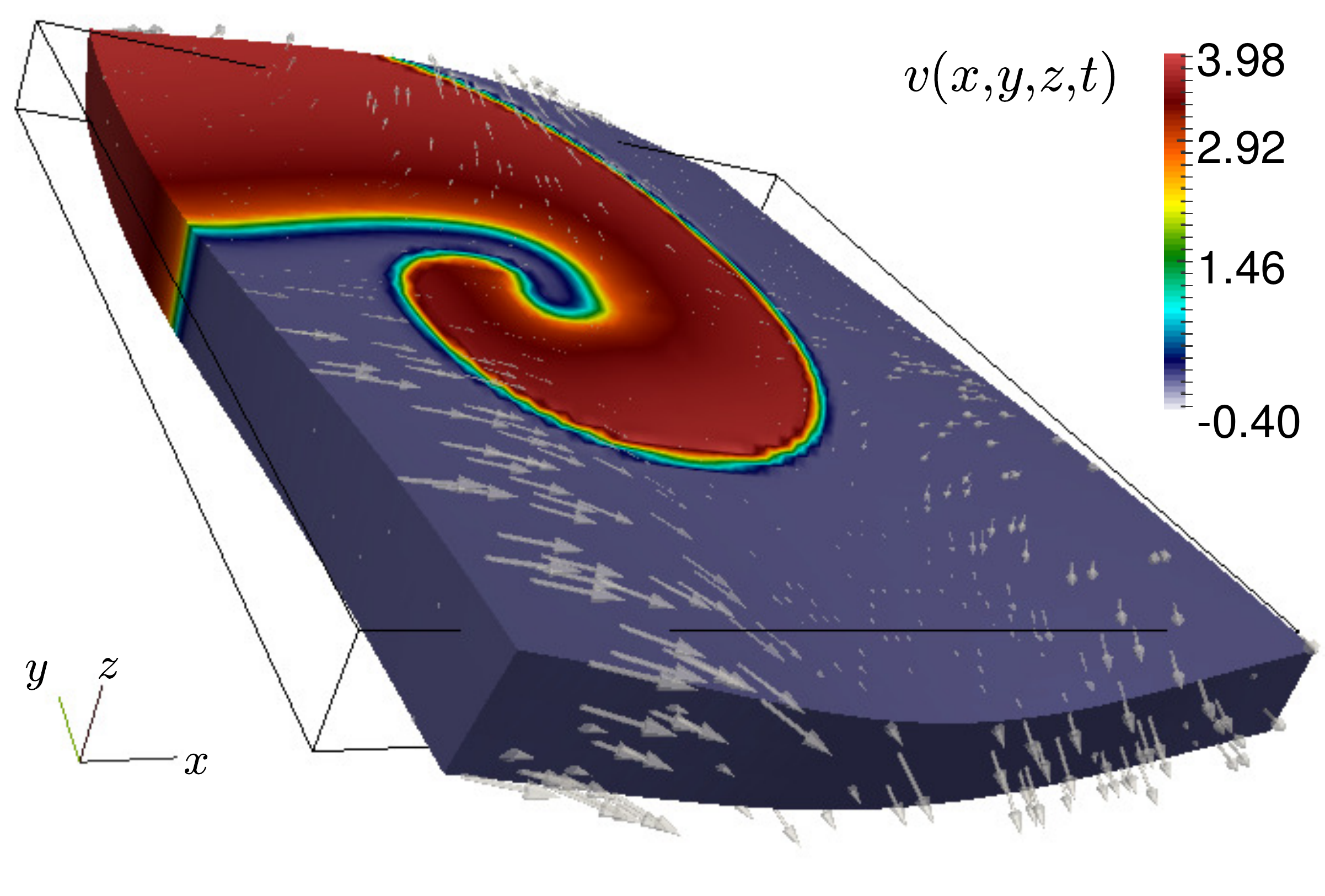}
\includegraphics[width=0.495\textwidth]{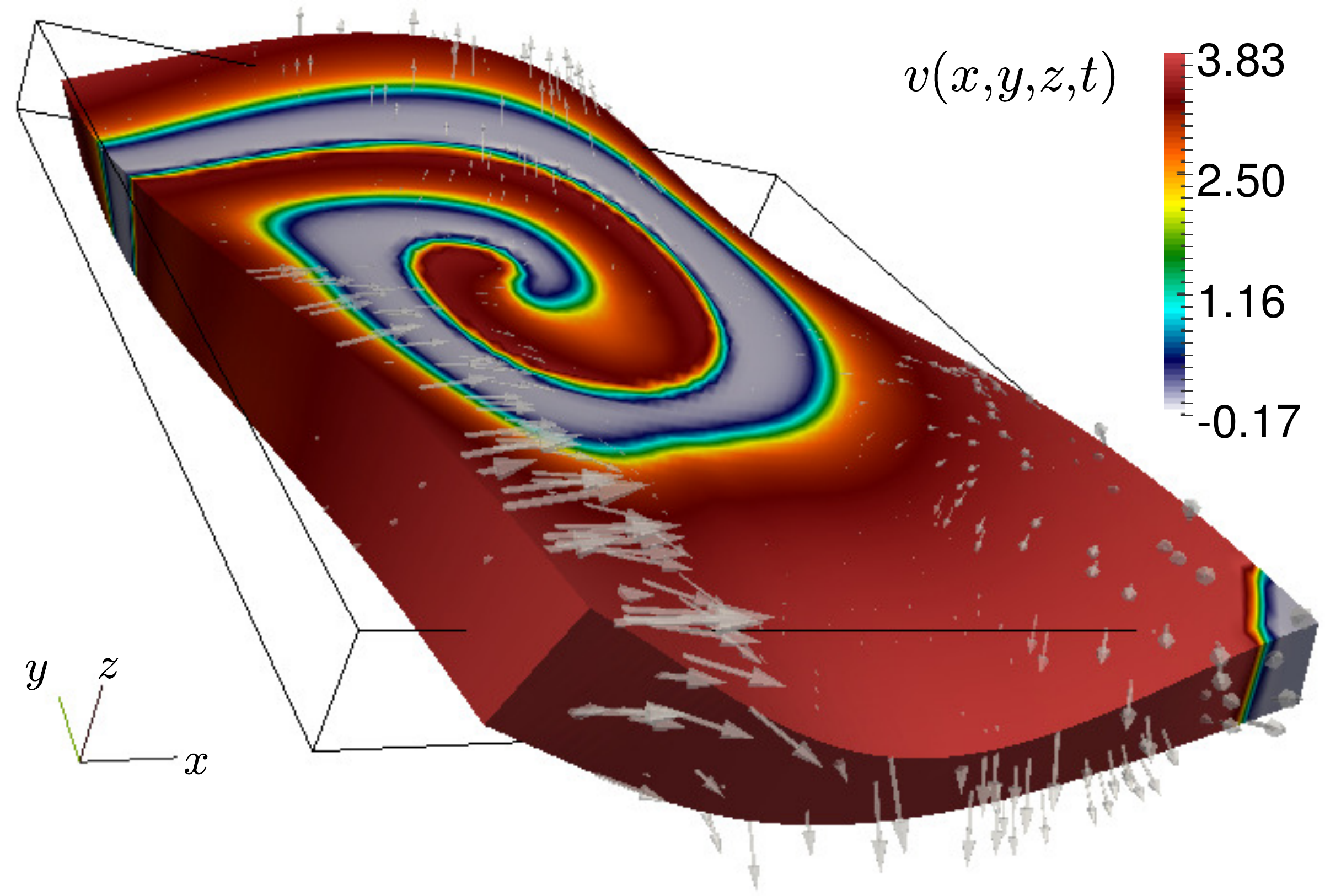}
\end{center}
\vspace{-0.5cm}
\caption{Scroll waves developed after $t=450$\,ms {(left panel)} and $t = 600$\,ms {(right panel)} using $T=37^\circ$C, 
plotted on the deformed configuration, and where arrows indicate displacement directions.}
\label{fig:ex02a}
\end{figure}

\begin{figure}[t!]
\begin{center}
\includegraphics[width=0.495\textwidth]{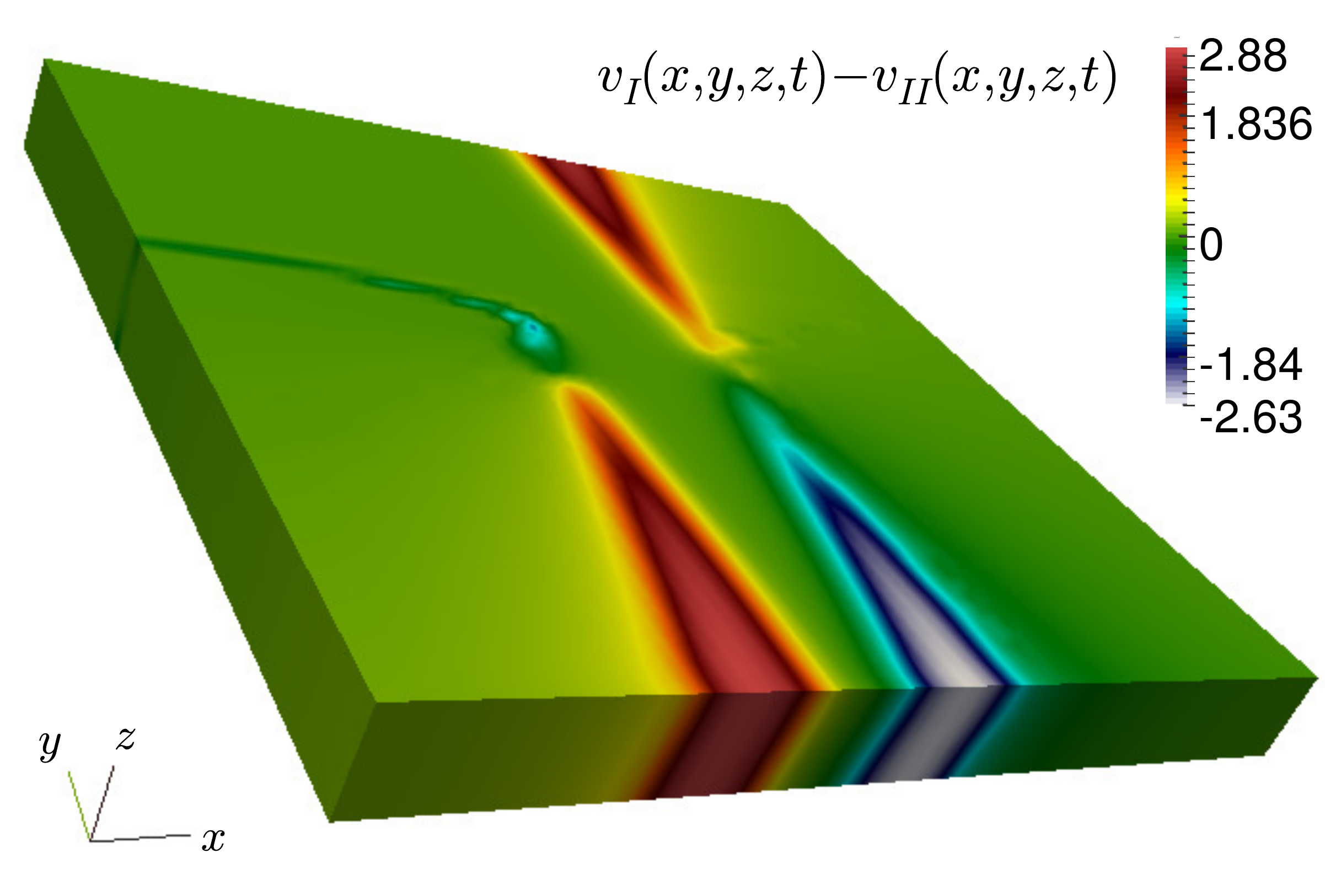}
\includegraphics[width=0.495\textwidth]{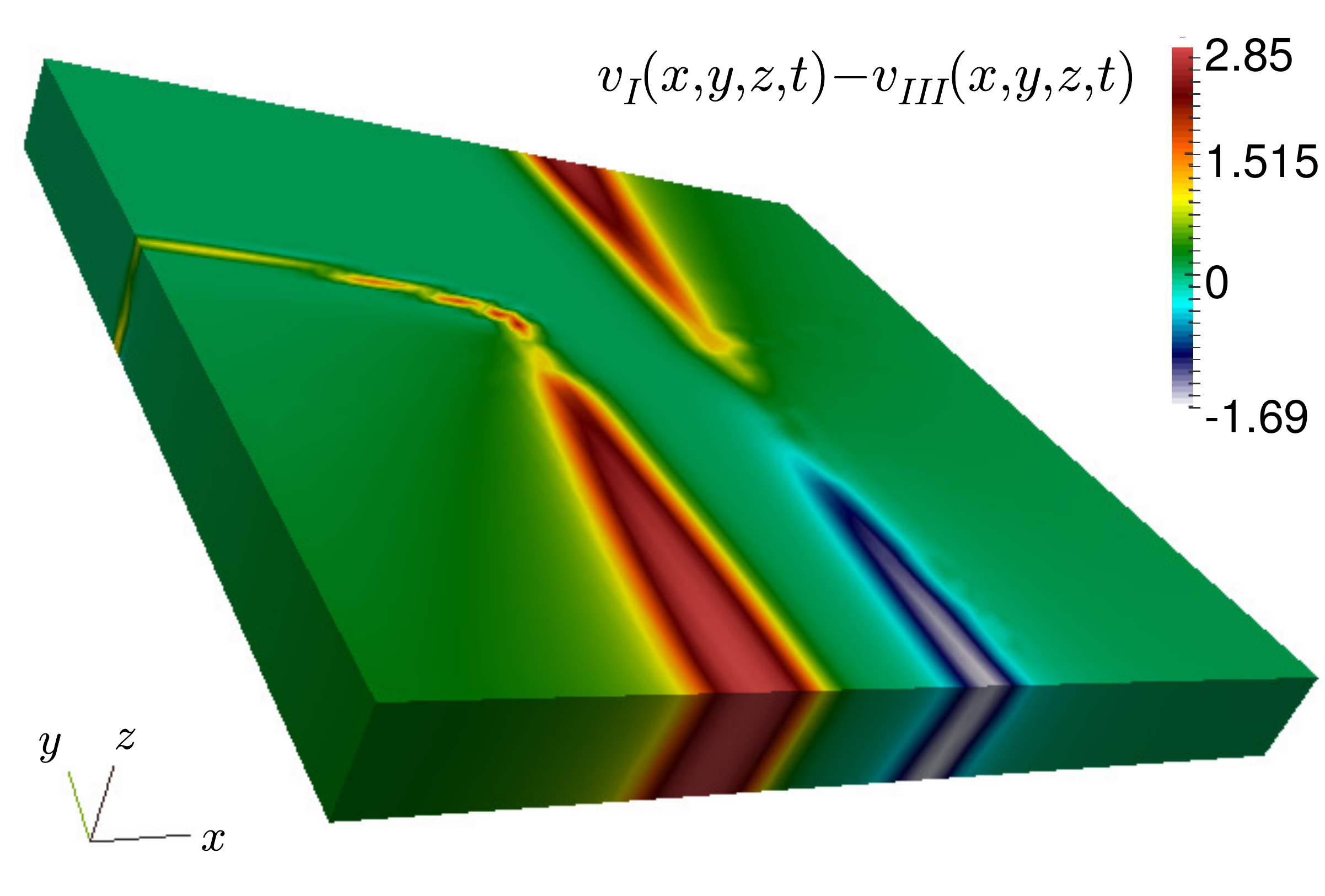}\\
\includegraphics[width=0.495\textwidth]{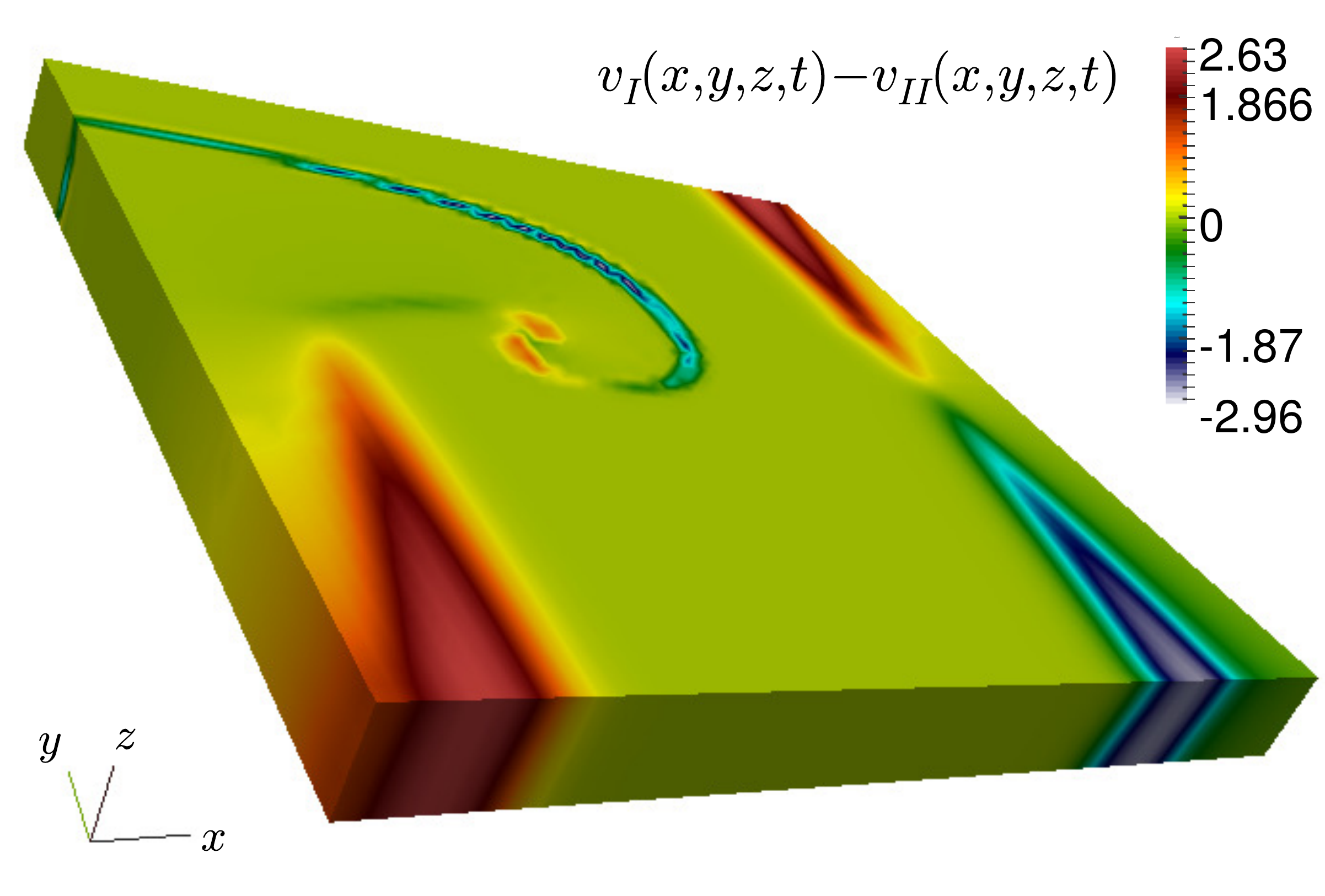}
\includegraphics[width=0.495\textwidth]{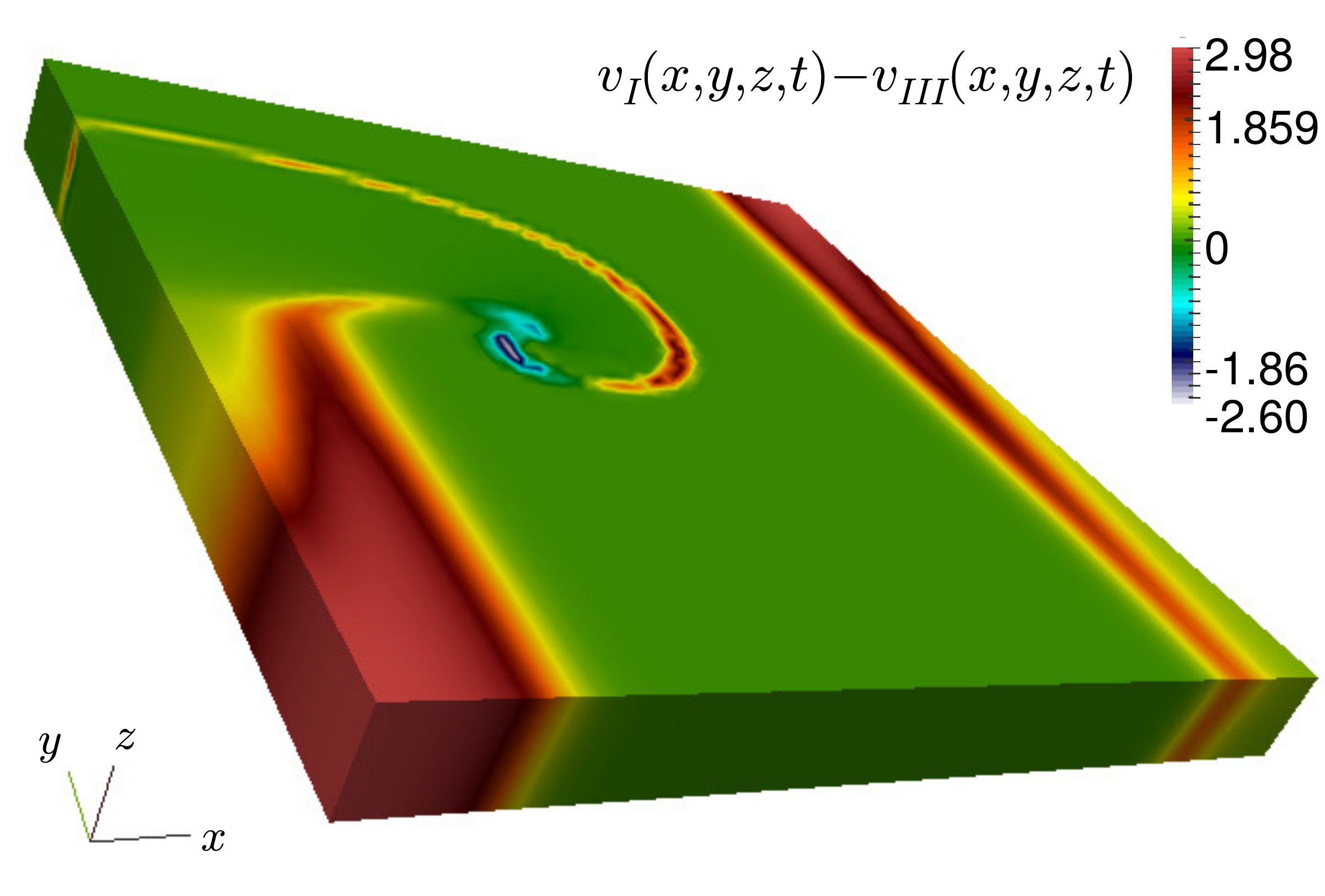}\\
\includegraphics[width=0.495\textwidth]{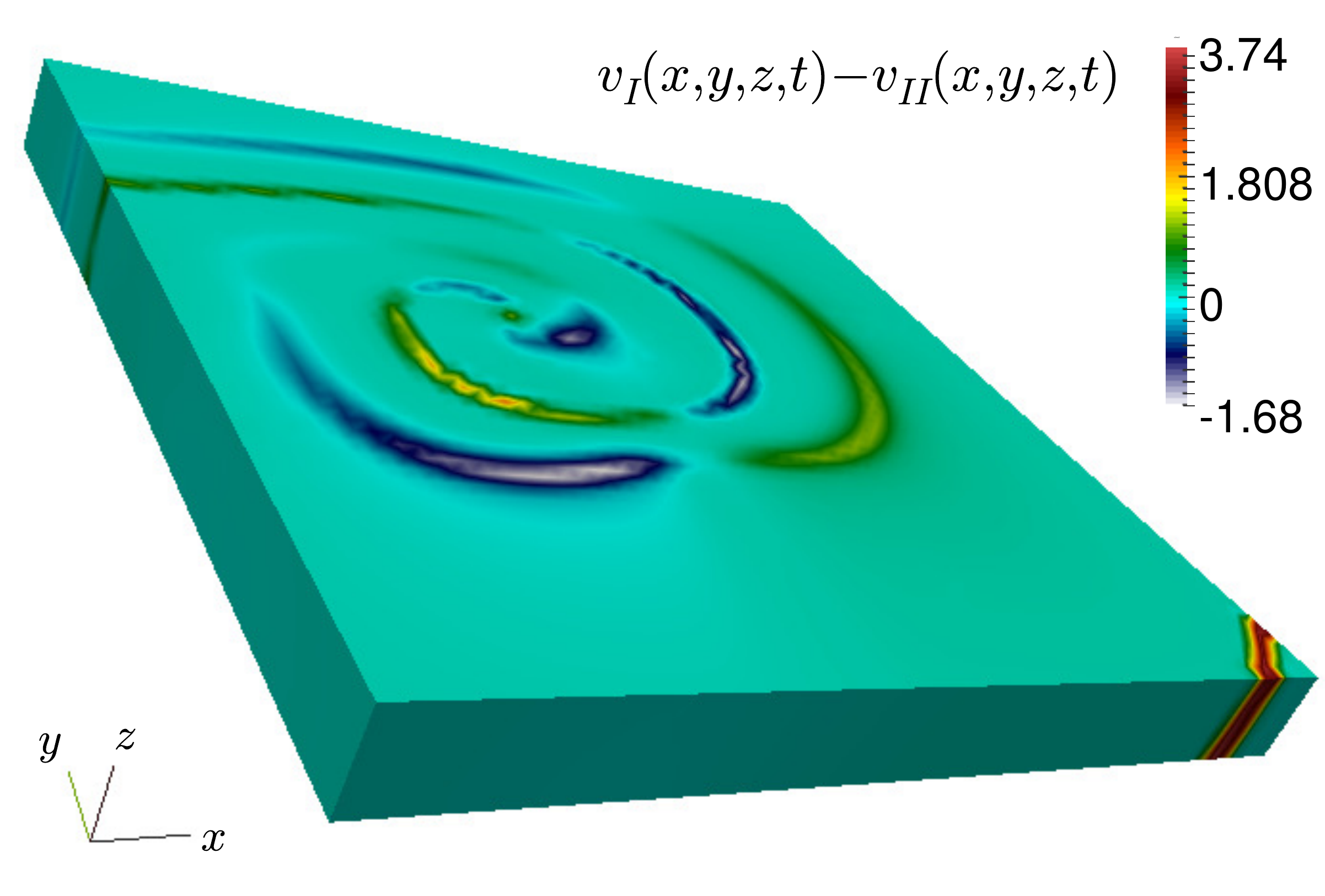}
\includegraphics[width=0.495\textwidth]{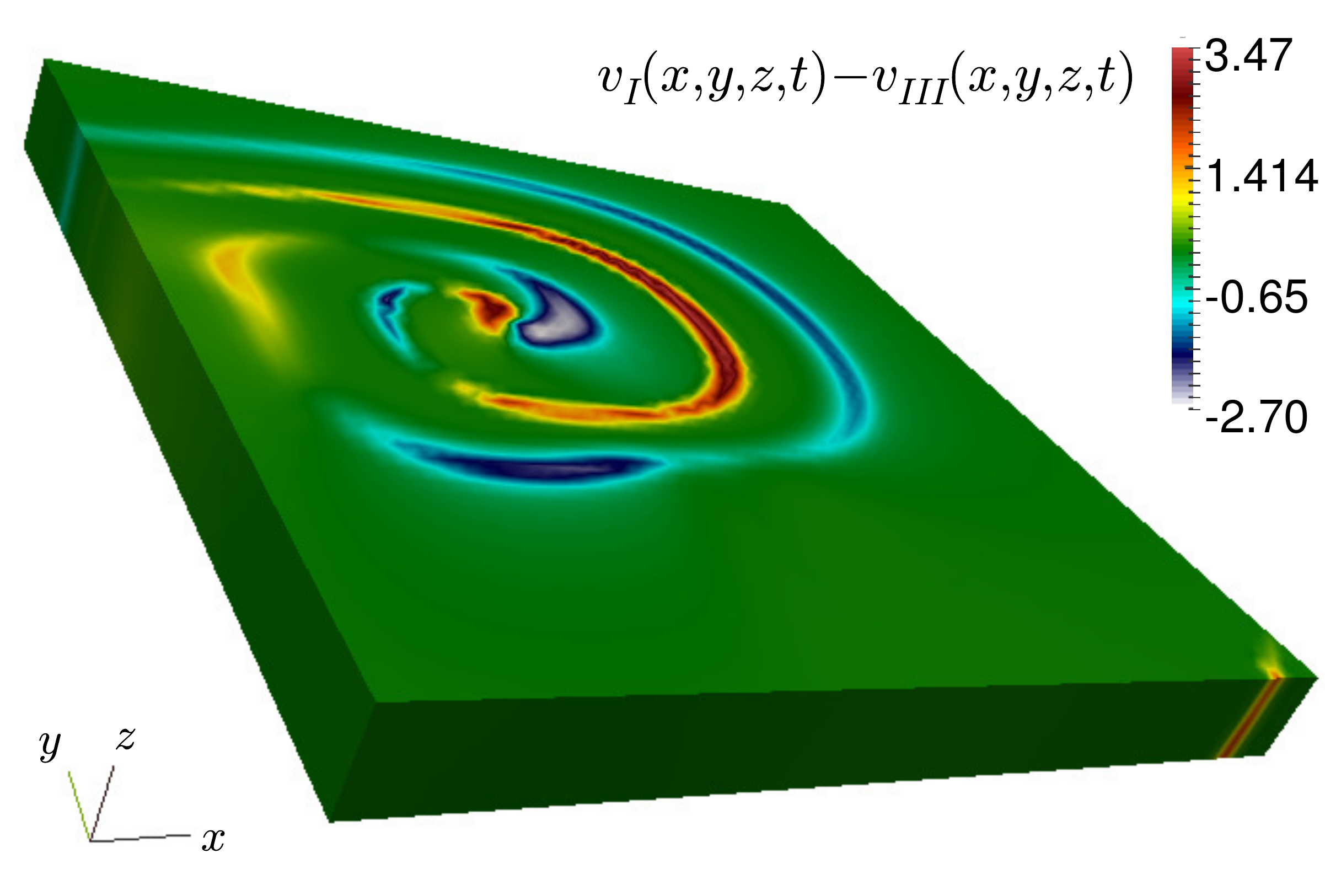}
\end{center}

\vspace{-0.5cm}
\caption{\cred{Potential difference between case I (uniform temperature 
at $T=37^\circ$C) and two different gradient distributions in the sheetlet 
(case II - left) and radial directions (case III - right), plotted on the reference domain
 at times $t=350$\,ms (top panels), $t=450$\,ms (middle row), and $t = 600$\,ms (bottom panels)}.}
\label{fig:ex02b}
\end{figure}

The domain of interest is now  
the slab $\Omega =(0,6.72)\times(0,6.72)\times(0,0.672)$\,{cm$^3$}, which we 
discretise into a structured mesh of 72'000 hexahedral elements, \cred{with $h=0.116$\,cm}.
We use a fixed timestep $\Delta t = 0.03\,\rm ms$ and set a constant 
fibre direction $\fo = (1,0,0)^{\tt t}$. We employ an S1-S2 protocol 
to initiate scroll waves~\citep{bini:2010}, where S1 is \cred{a  square} wave stimulation current of amplitude 
\cred{3} and duration 3\, ms, 
starting at $t=1$\, ms on the face defined by $x=0$; and 
S2 is a step function of the same duration and amplitude, 
applied on the lower left octant of the domain at $t=350\,$ ms. This time  
the boundary conditions for the structural problem are precisely as in 
\eqref{eq:robin}, using the constant $\eta = 0.05$; and the boundary 
conditions for the electrophysiology adopt the form \eqref{eq:zeroflux}. 
Figure \ref{fig:ex02a} shows two snapshots of the voltage propagation 
through the deformed tissue slab for the first case, of constant 
temperature (case I). Differences  between 
the patterns obtained at different temperatures are qualitatively shown 
in Figure~\ref{fig:ex02b}, which displays the difference in the potential 
between case II and case I, as well as between case III and case I. 
A fourth case (not shown) was also tested, where the temperature gradient is 
placed in the direction of the fibres. Then the differences in 
propagation are much more pronounced (up to the point that the 
S1-S2 protocol described above is not able to produce scroll waves).

\begin{figure}[t!]
\begin{center}
\includegraphics[width=0.325\textwidth]{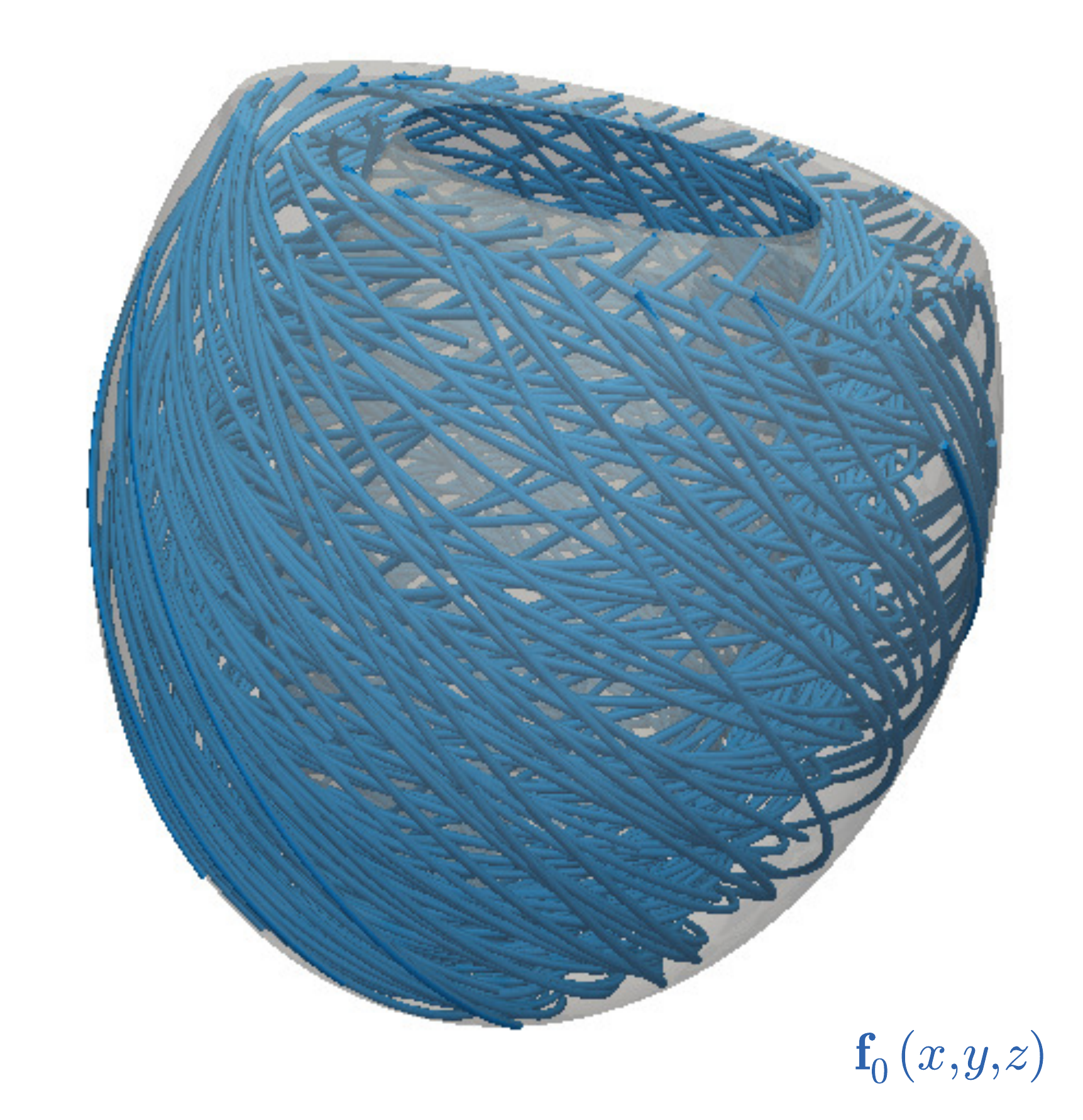}
\includegraphics[width=0.325\textwidth]{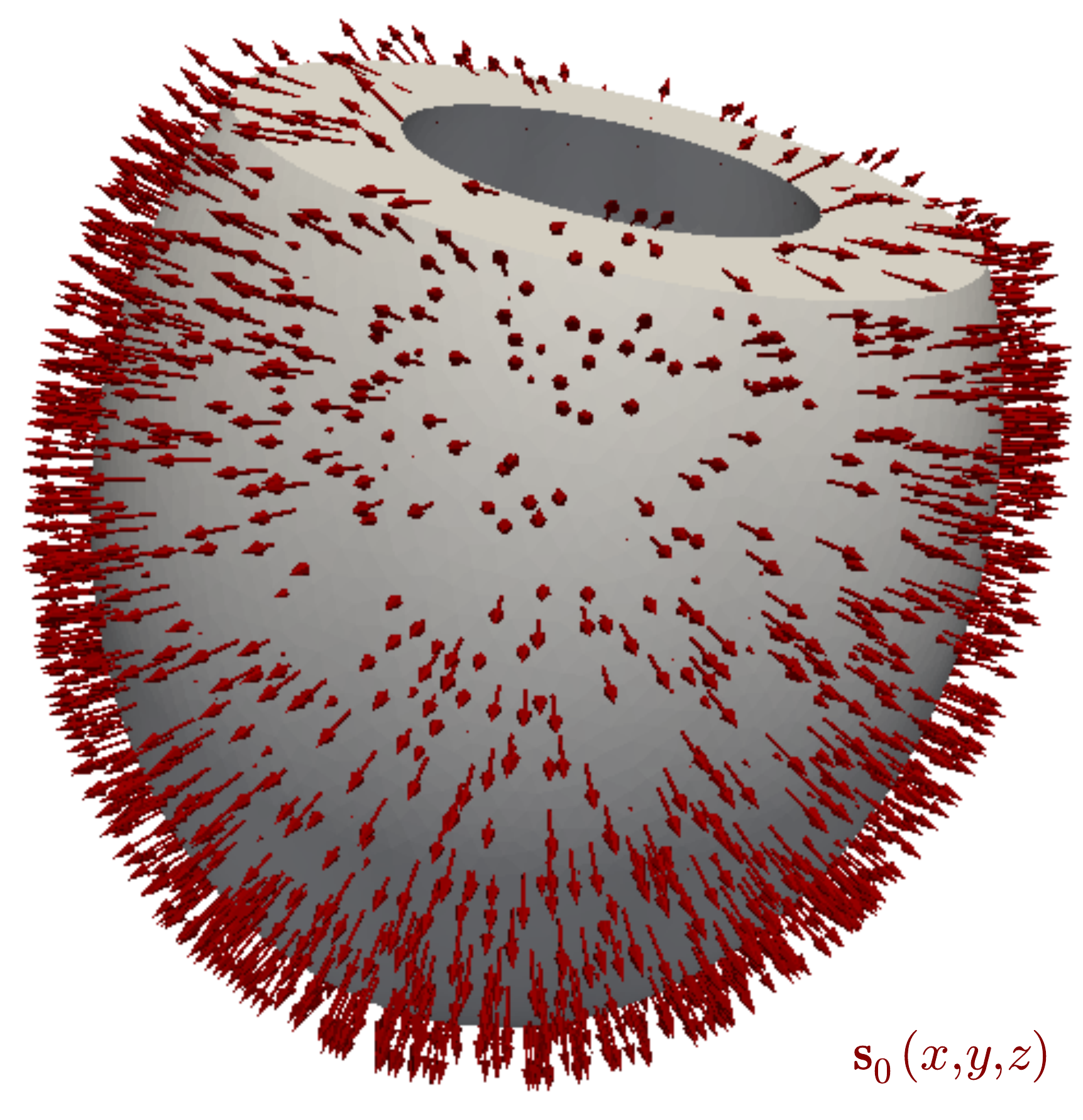}
\includegraphics[width=0.325\textwidth]{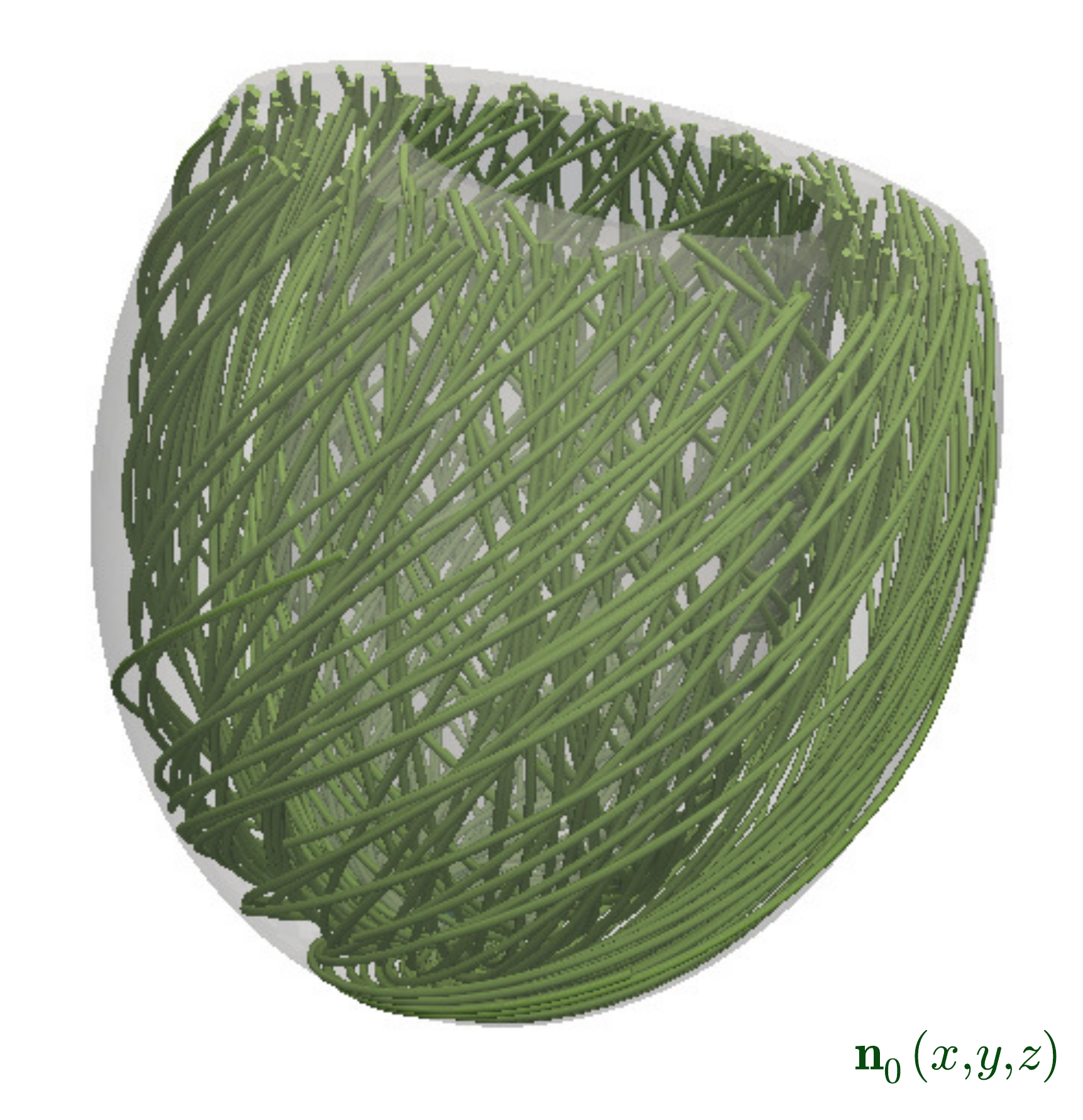}
\end{center}

\vspace{-0.5cm}
\caption{\cred{Ellipsoidal fibre distribution, collagen normal-sheetlet, and cross-fibre directions generated with a 
rule-based algorithm and setting} $\theta_{\text{epi}}=-50^\circ$, 
$\theta_{\text{endo}}=60^\circ$.}
\label{fig:fibres}
\end{figure}

\subsection{Scroll waves in an idealised left-ventricular geometry}\label{sec:ellipsoid}
We generate the geometry of a truncated ellipsoid, as well as unstructured 
hexahedral meshes using GMSH \citep{gmsh}. The domain has a height (base-to-apex) 
of 6.8\,cm, a maximal equatorial diameter of 6.6\,cm, a ventricular 
thickness of 0.5\,cm at the apex and of 1.3\,cm at the equator. Relatively coarse and fine partitions 
with 13'793 \cred{(corresponding to a meshsize of $h=0.104$\,cm) and 86'264 elements (and with 
a meshsize of $h=0.052$\,cm)} are used for the simulations in this subsection. Consistently with 
other electromechanical simulations on idealised ventricular geometries, here we consider 
a time-dependent pressure distributed uniformly on the endocardium (that is, using 
the second relation in \eqref{eq:mixedBC}). In addition, on the basal cut we impose 
zero normal displacements (the first condition in \eqref{eq:mixedBC}), and on the 
epicardium we impose Robin conditions \eqref{eq:robin} setting a spatially varying
stiffness coefficient, going linearly from \cred{$\eta_{\min}$ on the 
apex, to $\eta_{\max}$} on the base  
$\eta(y):=\frac{1}{y_b-y_a} [\eta_{\max}\cred{(y_b-y)} + \eta_{\min}(y - y_a)]$, 
where $y_a,y_b$ denote the vertical component of the positions at the apex and base, 
respectively. {These conditions are sufficiently general to mimic} the presence of the pericardial sac (as well as 
the combined elastic effect of other surrounding organs) {having spatially-varying stiffness.}

\begin{figure}[t!]
\begin{center}
\raisebox{3mm}{\includegraphics[width=0.29\textwidth]{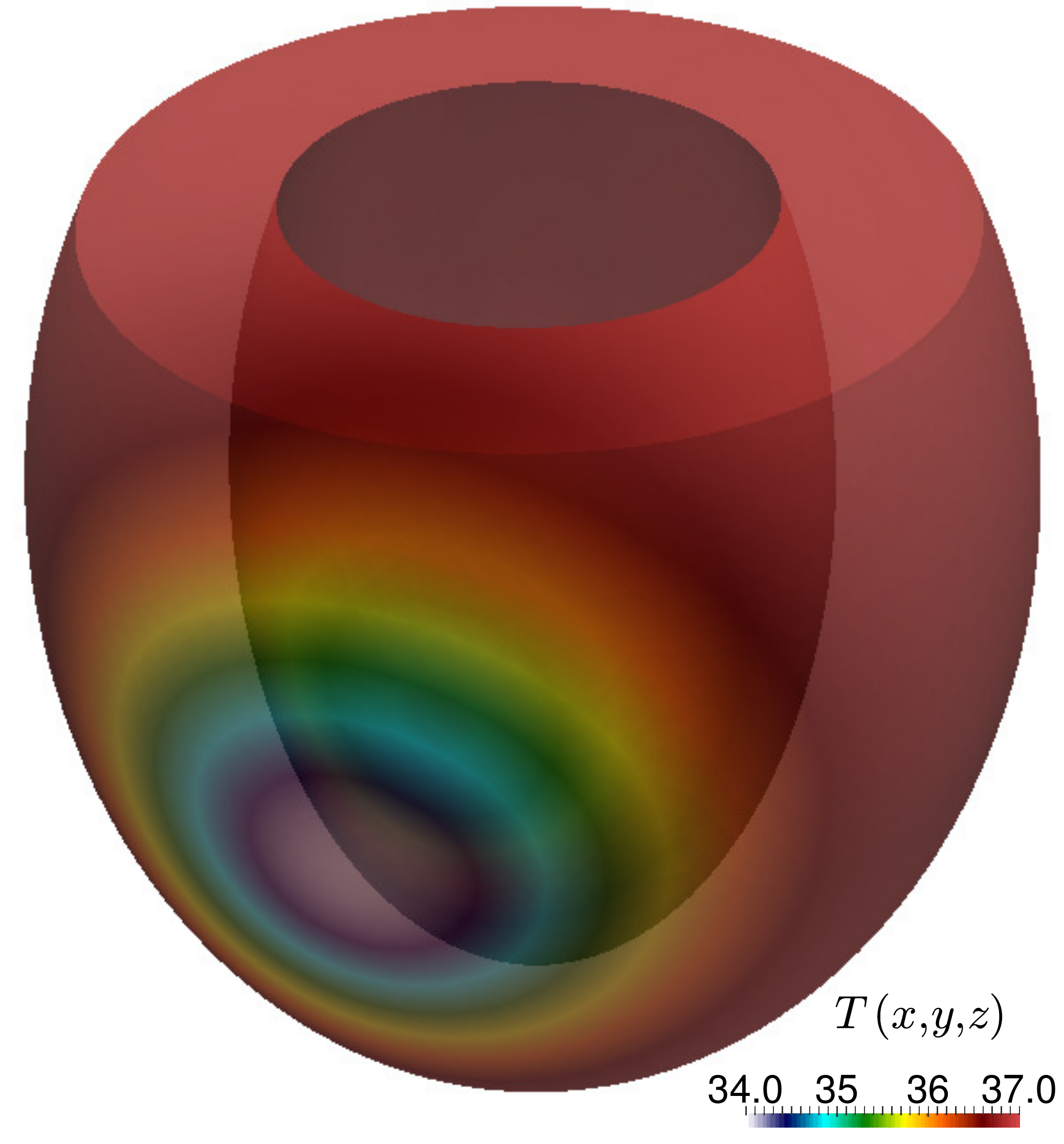}}\quad
\includegraphics[width=0.325\textwidth]{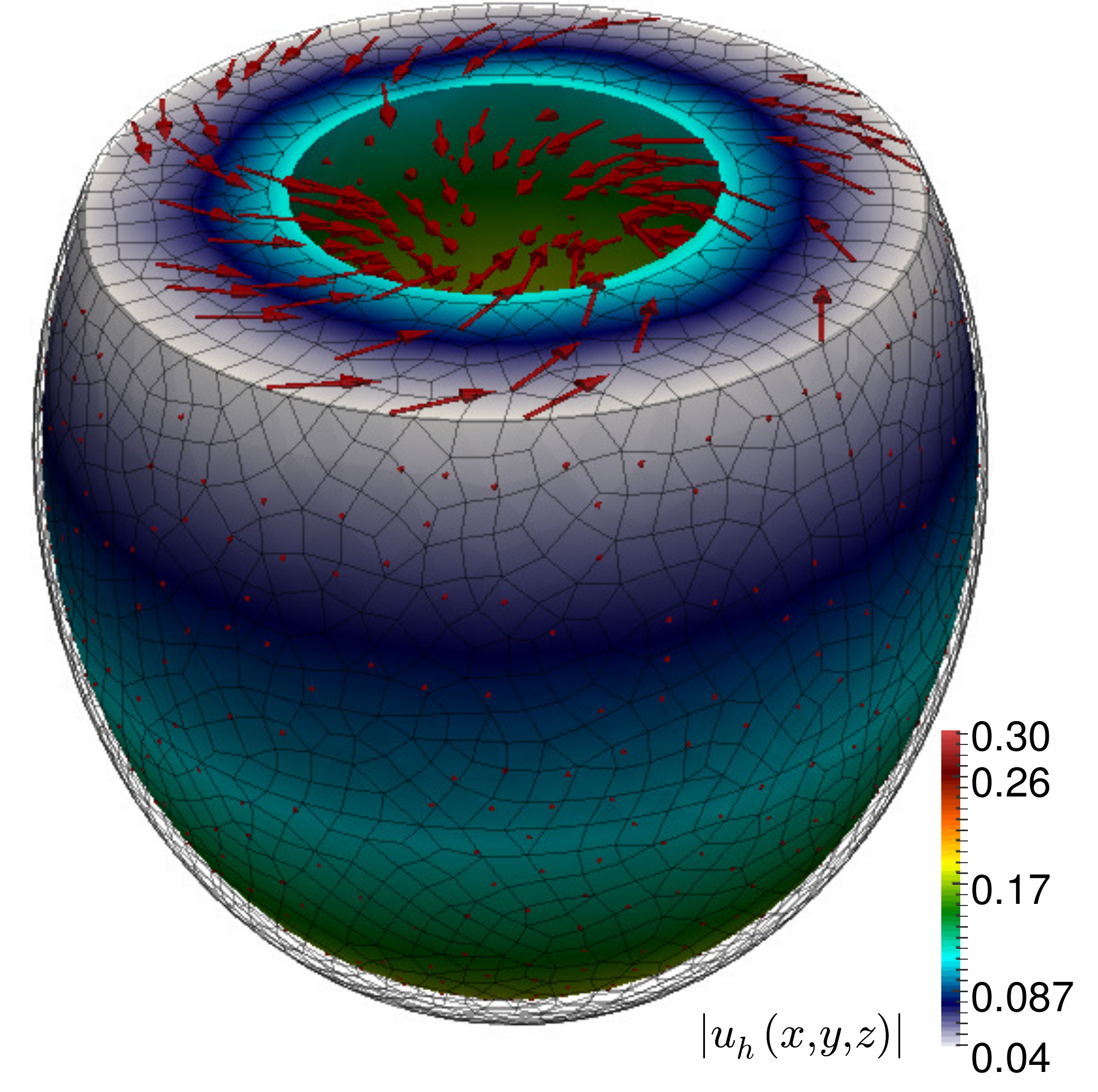}
\includegraphics[width=0.325\textwidth]{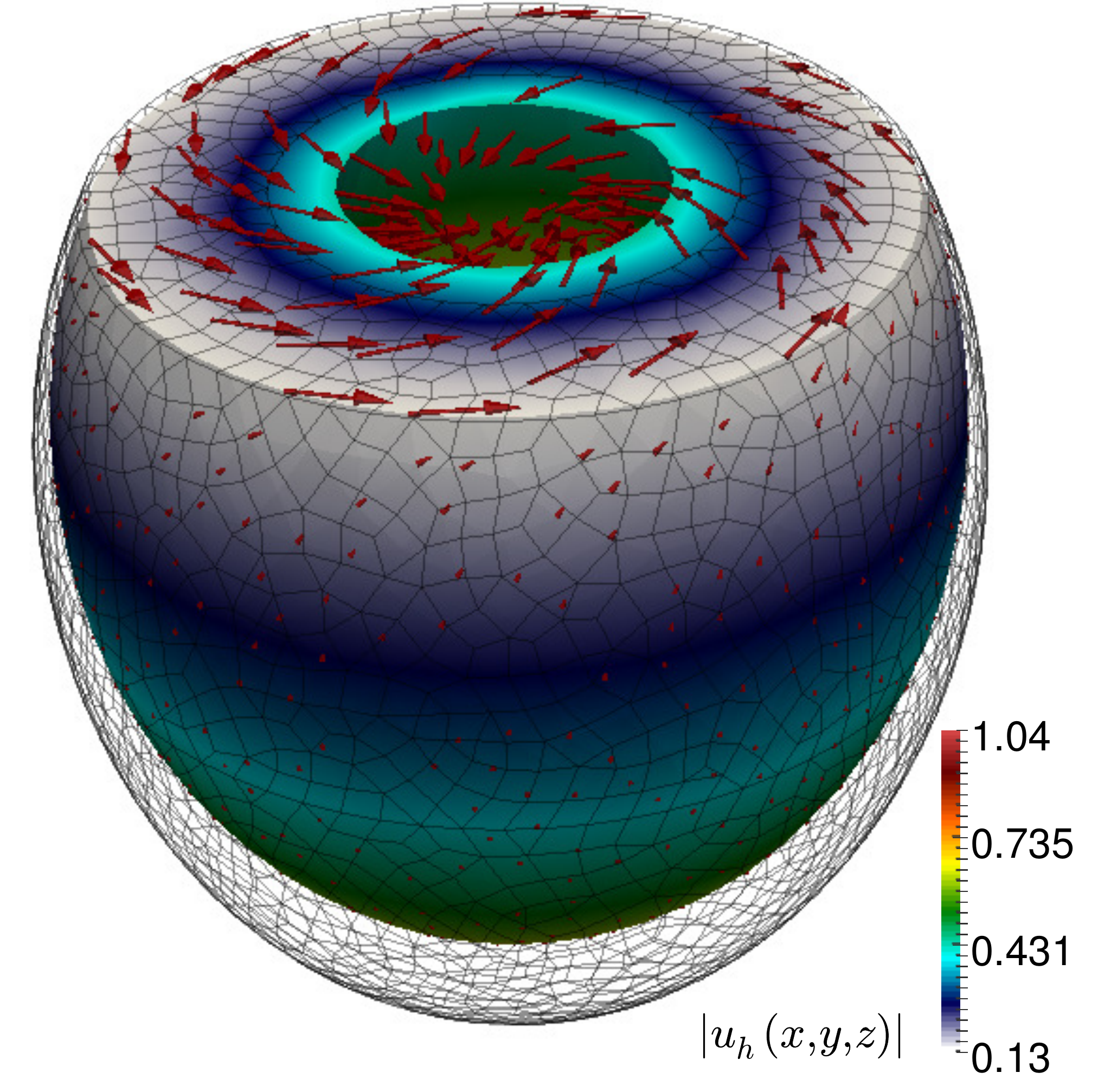}\\
\includegraphics[width=0.4\textwidth]{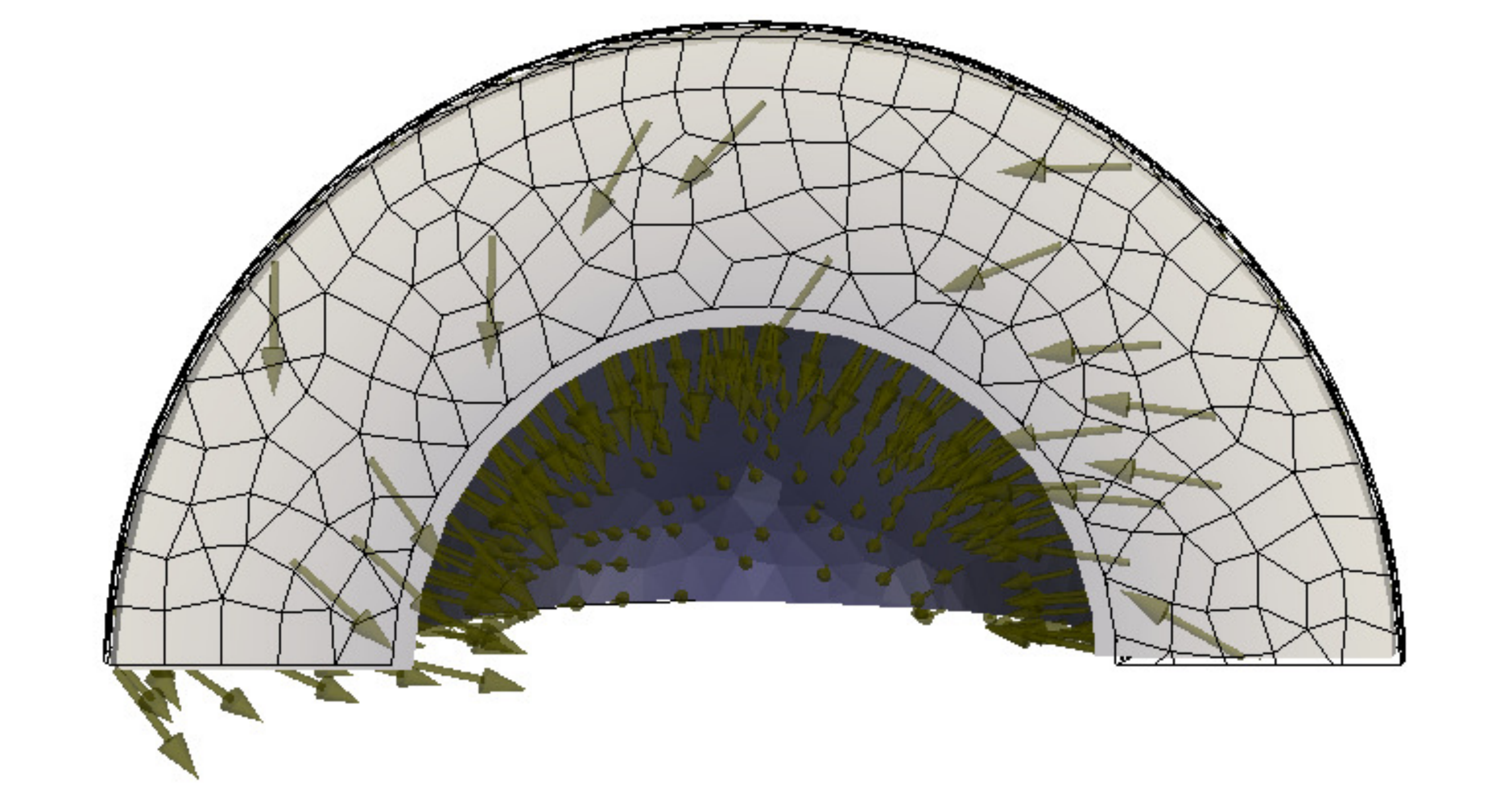}\qquad
\includegraphics[width=0.4\textwidth]{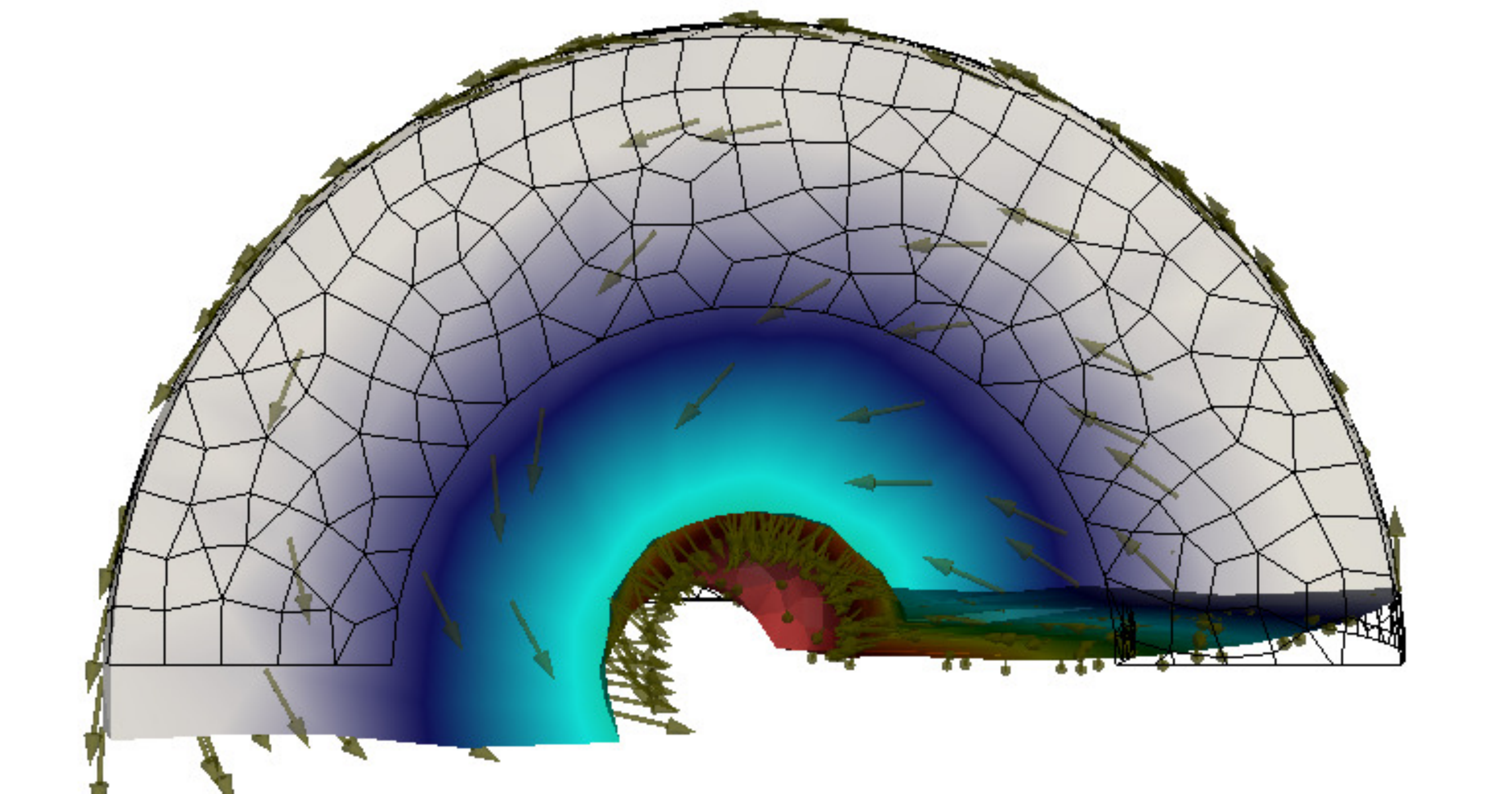}
\end{center}

\vspace{-0.5cm}
\caption{Temperature distribution for the second test case (top left), and two 
snapshots (200,300\,ms after the S1 stimulus) illustrating the torsion and wall-thickening of the left ventricle (top centre and 
top right, where the arrows indicate 
the displacement direction). 
The bottom panels show cuts along the $z-$midplane for the 
 top centre and top right figures.}
\label{fig:torsion}
\end{figure}

\begin{figure}[t!]
\begin{center}
\includegraphics[width=0.245\textwidth]{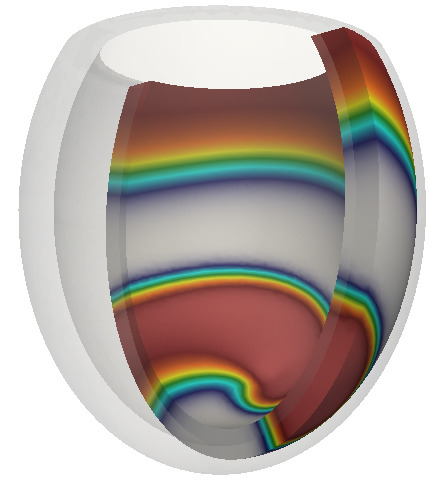}
\includegraphics[width=0.245\textwidth]{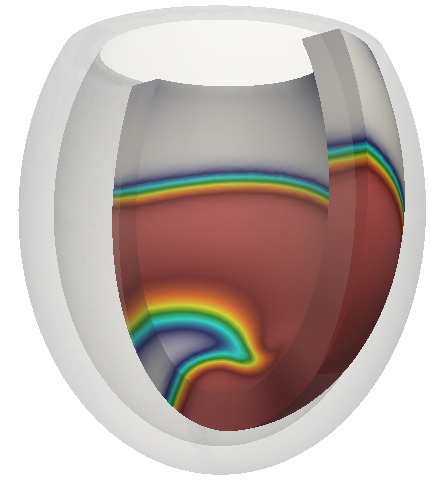}
\includegraphics[width=0.245\textwidth]{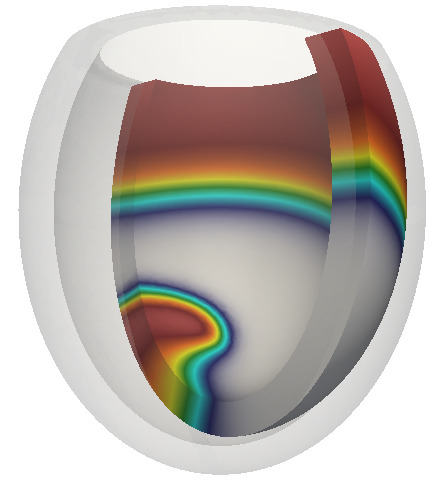}
\includegraphics[width=0.245\textwidth]{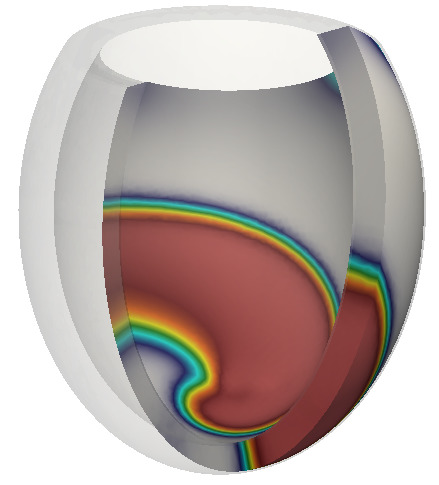}\\
\includegraphics[width=0.245\textwidth]{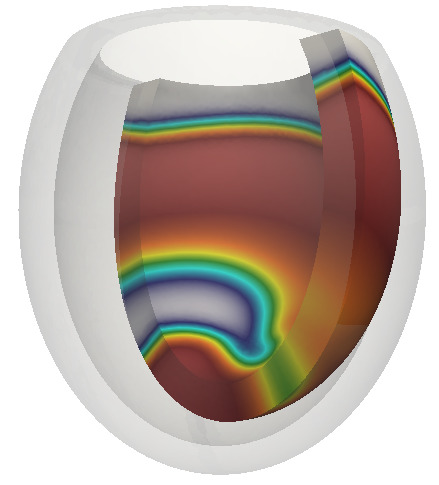}
\includegraphics[width=0.245\textwidth]{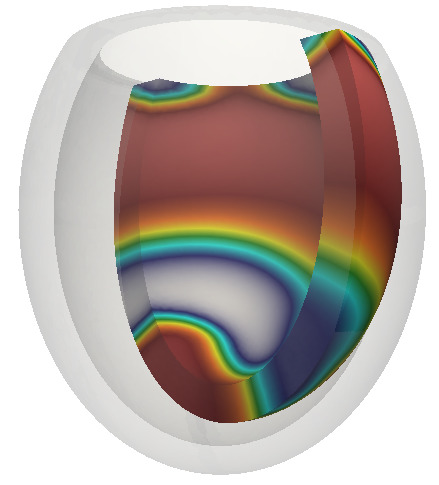}
\includegraphics[width=0.245\textwidth]{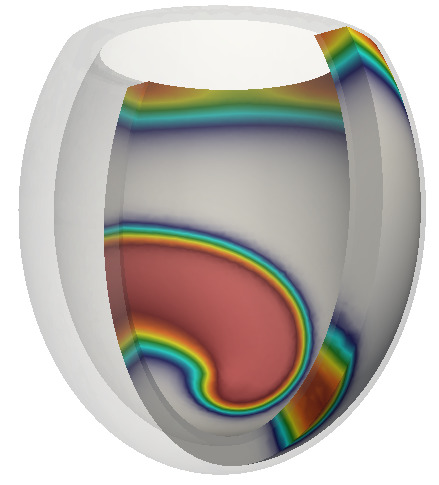}
\includegraphics[width=0.245\textwidth]{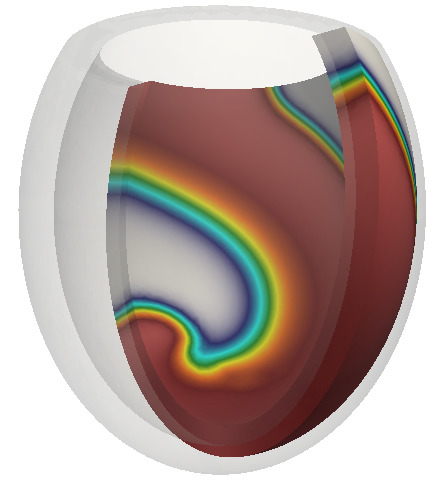}\\
\includegraphics[width=0.35\textwidth]{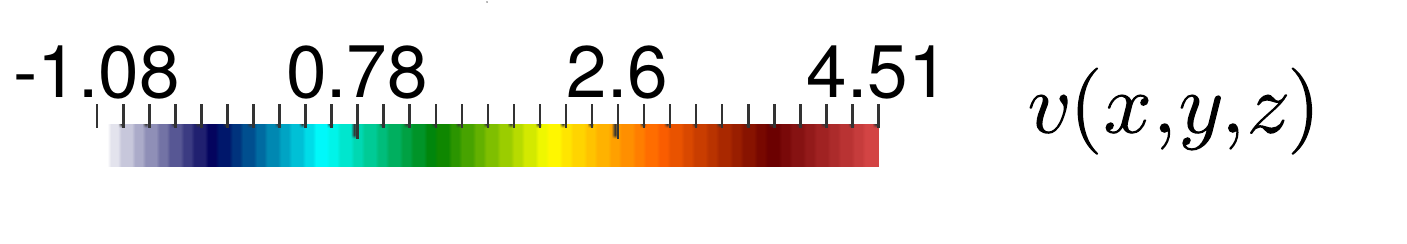}
\end{center}

\vspace{-0.5cm}
\caption{Propagation of the transmembrane potential plotted 
on the deformed domain, using a constant temperature (top panels) and a 
cold spot (bottom). \cred{Snapshots shown at 200,300,400,500\,ms after the S2 stimulus.}}
\label{fig:ellip}
\end{figure}

Fibre and sheetlet directions are constructed using a slight modification to the 
rule-based algorithm proposed in \cite{rossi:2014}, that we outline here for the sake of 
completeness (see Algorithm~\ref{alg:EM}).  The needed inputs are a unit vector 
$\ko$ aligned with the centreline and pointing from apex to base, the desired maximal 
and minimal angles that will determine the rotational anisotropy from epicardium to 
endocardium, $\theta_{\text{epi}}$, $\theta_{\text{endo}}$; and boundary labels for 
the epicardium $\partial\Omega_{\text{epi}}$, endocardium  $\partial\Omega_{\text{endo}}$, 
and basal cut  $\partial\Omega_{\text{base}}$. The first step consists in solving the following Poisson 
problem (here stated in mixed form for a potential $\phi$ and a preliminary sheetlet direction  $\bzeta$) 
endowed with mixed boundary conditions 
\begin{equation}\label{eq:problem-fibres}
\begin{split}
- \nabla \cdot \bzeta & = 0 \quad\text{and}\quad \bzeta = \nabla \phi \qquad \text{in $\Omega$},\\
\bzeta \cdot \bn & = 0 \quad \text{on } \partial\Omega_{\text{base}}, \quad 
\phi  = 0 \quad \text{on } \partial\Omega_{\text{endo}}, \quad \phi = 1 \quad 
\text{on } \partial\Omega_{\text{epi}}.
\end{split}\end{equation}
The unknowns of this problem are discretised with Brezzi-Douglas-Marini 
elements of first order {defined on quads}, and piecewise constant elements \citep{gatica:book}. Once a 
discrete first sheetlet direction $\bzeta_h$ is computed, the final sheetlet 
directions are obtained by normalisation $\so = \bzeta_h / \|\bzeta_h\|$ \cred{(all 
 normalisations in this {section} refer to component-wise operations using the Euclidean norm)}.
Secondly, we project the centreline $\widehat{\ko} = \ko - (\ko \cdot \so) \so$   
and then 
compute \cred{an auxiliary vector field $\widehat{\fo}$ (known as flat fibre field)}, 
using the sheetlet and the projected centreline 
vectors $\widehat{\fo} = \so \times \widehat{\ko}/\|\widehat{\ko}\|$. 
Thirdly, we proceed to project now the flat fibres onto the sheetlet planes exploiting 
the rotational anisotropy, through the  operation 
 \[\cred{\fo} = \widehat{\fo} \cos( \theta(\phi_h)) + \so \times \widehat{\fo} \sin(\theta(\phi_h)) 
 + \so (\so\cdot \widehat{\fo}) [1 - \cos(\theta(\phi_h))],\] 
 where $\phi_h$ is the 
 discrete potential and the function 
 $$\theta(\phi_h):=\frac{1}{180\pi} [(\theta_{\text{epi}} - \theta_{\text{endo}} ) \phi_h 
 + \theta_{\text{endo}}],$$ modulates the intramural angle variation. 
 Sample 
 fibre, sheet and normal directions generated using this algorithm are shown in 
Figure~\ref{fig:fibres}.

The remaining constants employed in this Section are $\theta_{\text{epi}}=-50^\circ$, 
$\theta_{\text{endo}}=60^\circ$, $\eta_{\max}=0.6\,$kPa (that is, we consider 
a transmurally asymmetric fibre distribution), and $\eta_{\min}=0.001\,$kPa. 
\cred{As in the tests reported in previous subsections, the dynamics here 
are initiated through an S1-S2 approach~\citep{karma:2013}, which is a standard 
 stimulation protocol in cardiac electrophysiology (both 
experimentally and \textit{in silico}), aimed at determining spiral wave inducibility, 
in the context of replicating archetypal features of cardiac arrhythmias.
One typically generates a planar electrical excitation (S1), followed by
a second broken stimulus (S2) during the repolarisation phase of the
S1 wave, the so-called vulnerable window. 
In our numerical simulations, S1 is set on the apex and S2 is initiated at the same location, 
but only for the quadrant $x>0,z>0$.}

\begin{figure}[t!]
\begin{center}
\includegraphics[width=0.245\textwidth]{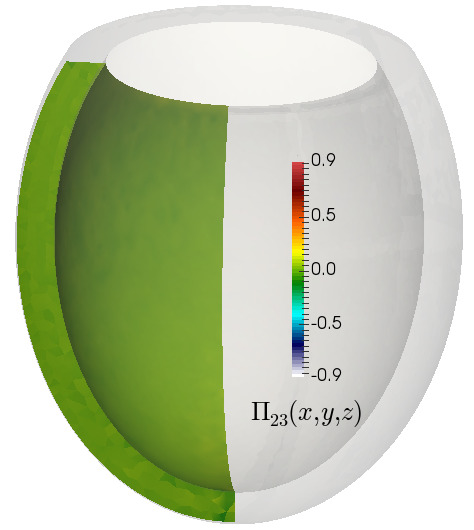}
\includegraphics[width=0.245\textwidth]{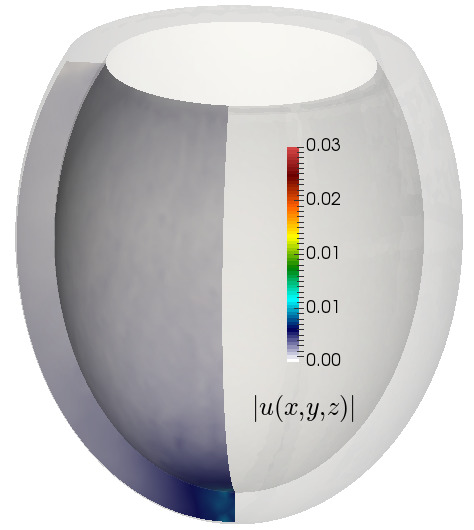}
\includegraphics[width=0.245\textwidth]{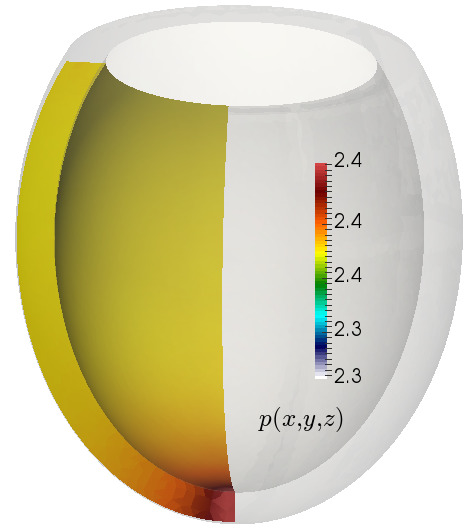}
\includegraphics[width=0.245\textwidth]{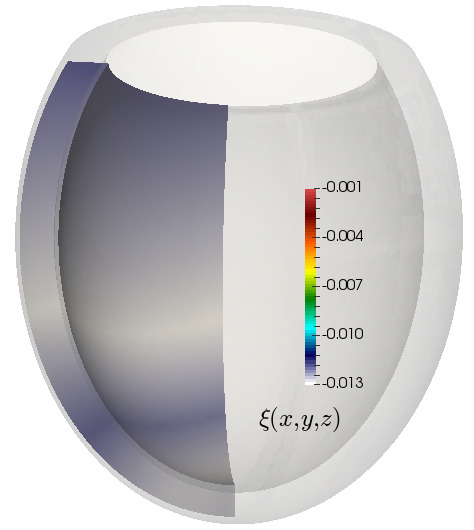}\\
\includegraphics[width=0.245\textwidth]{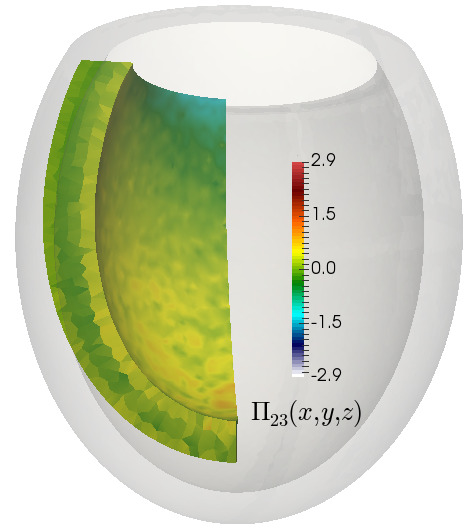}
\includegraphics[width=0.245\textwidth]{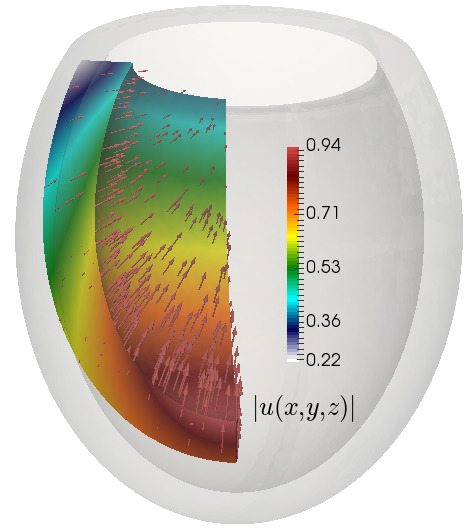}
\includegraphics[width=0.245\textwidth]{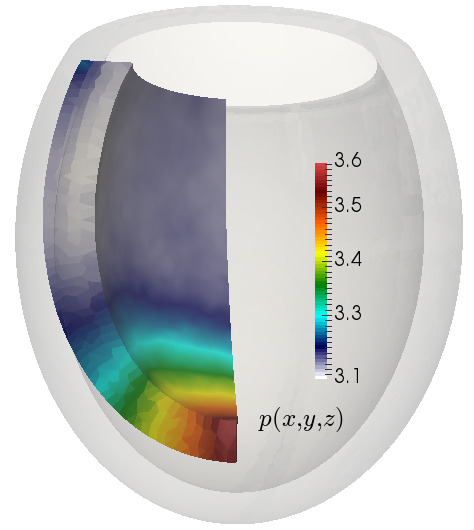}
\includegraphics[width=0.245\textwidth]{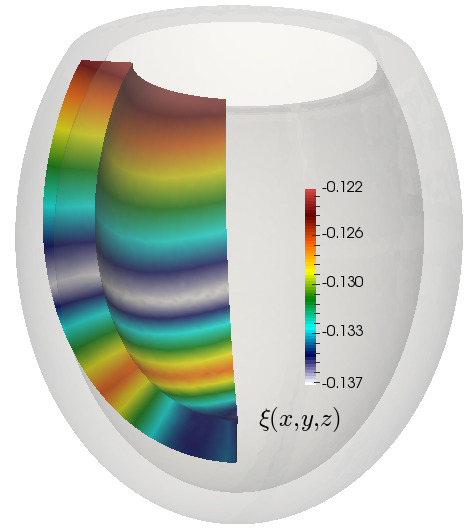}
\end{center}

\vspace{-0.5cm}
\caption{\cred{Sample approximate Kirchhoff stress, displacement, 
pressure, and myocyte shortening at end diastole, $t = $470\,ms (top) and 
$t = $610\,ms (bottom).}}
\label{fig:pi-u-p}
\end{figure}

We consider two cases: one when the 
temperature is kept constant at $37^\circ$C, and another when at the time 
of switching on the electromechanical coupling, a localised point {on the 
epicardium towards the base} is maintained at a lower temperature $34^\circ$C. The temperature 
distribution in this second case is defined as 
$$T(x,y,z) = 37 - 3\exp(-[(x-3)^3+y^2+z^2]/3),$$ (see the leftmost panel in Figure~\ref{fig:torsion}). 
We illustrate the torsion and wall-thickening effects achieved by the 
orthotropic activation model in the centre and right panels of Figure~\ref{fig:torsion}, \cred{observed 
before applying the wave S2}.

\begin{figure}[t!]
\begin{center}
\includegraphics[width=0.245\textwidth]{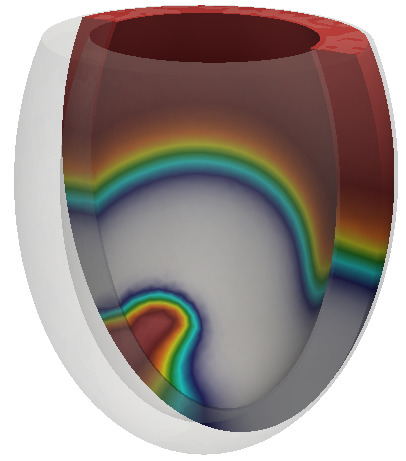}
\includegraphics[width=0.245\textwidth]{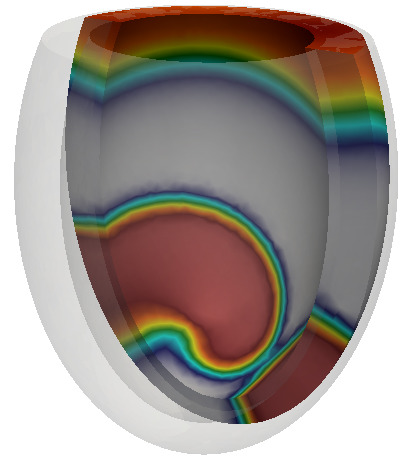}
\includegraphics[width=0.245\textwidth]{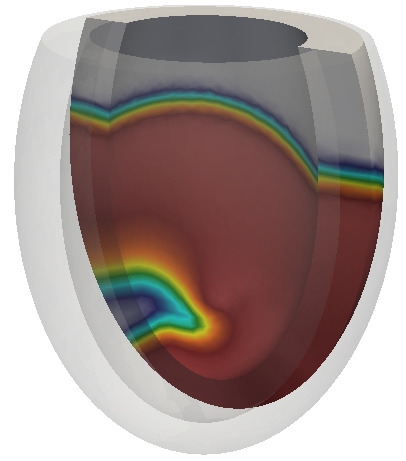}
\includegraphics[width=0.245\textwidth]{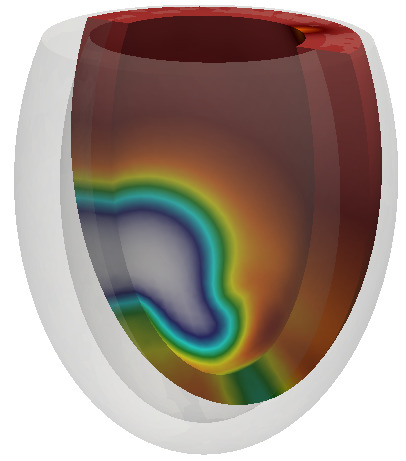}\\
\includegraphics[width=0.245\textwidth]{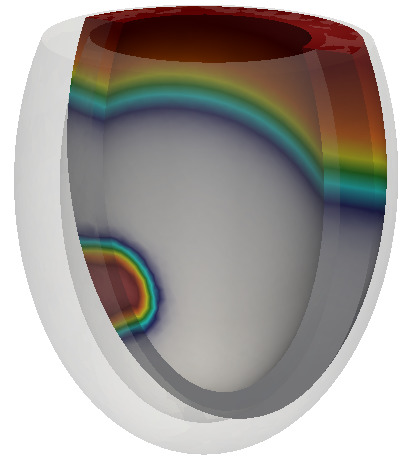}
\includegraphics[width=0.245\textwidth]{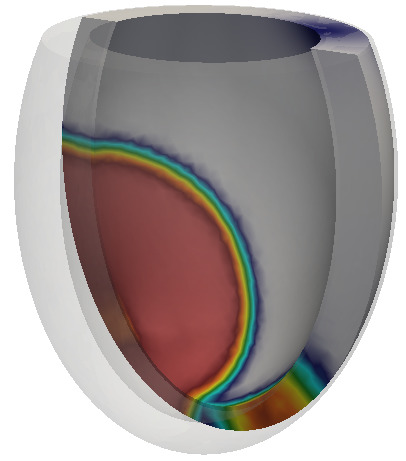}
\includegraphics[width=0.245\textwidth]{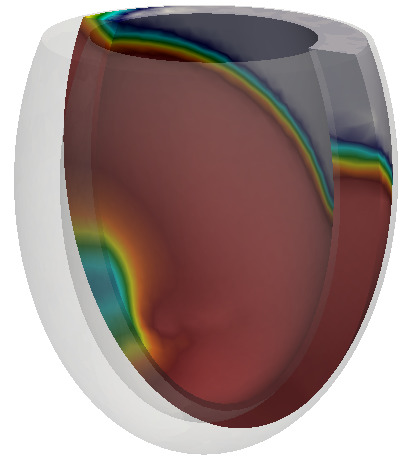}
\includegraphics[width=0.245\textwidth]{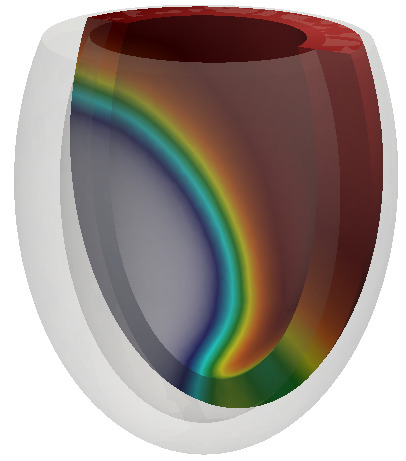}\\
\includegraphics[width=0.35\textwidth]{ex06color-crop}
\end{center}

\vspace{-0.5cm}
\caption{\cred{Propagation of the transmembrane potential plotted 
on the deformed domain, using a higher temperature throughout the domain and a thiner ventricular 
geometry. 
Snapshots shown at 100,200,\ldots,800\,ms after the S2 stimulus.}}
\label{fig:hot}
\end{figure}

Finally, a few snapshots of the  scroll wave dynamics for the two cases are presented in 
Figure~\ref{fig:ellip}, indicating again an important model dependency on 
temperature variations. 
\cred{In particular, the cold region notably increases the action potential
duration. 
Once the arrhythmic pattern is fully established, the differences between the two cases are increased 
since higher nonlinearities appear. Samples of stress entries, displacement, pressure, and 
myocyte contraction are in presented in Figure~\ref{fig:pi-u-p}, plotted on wedges that highlight ventricular 
thickening, stress concentrations on the endocardium, a more pronounced pressure profile near the apex,
 and apex to base motion. In addition to this test, we perform a set of simulations using a constant 
higher temperature at $39^\circ$C, and snapshots of the approximate potential at various time steps 
are displayed in Figure~\ref{fig:hot}. }

\section{Concluding remarks}\label{sec:concl}
We have advanced a new theoretical framework for the modelling of cardiac
electromechanics that incorporates active strain, anisotropic and nonlinear diffusion,  
and thermo-electrical coupling as main ingredients. 
\cred{The proposed models couple different multi-field and multi-scale (cell and sub-cell levels)  
phenomena, and they constitute a natural extension of porous medium electrophysiology \cite{hurtado:2016} 
to the case of cardiac electromechanics. The continuum homogenised approach features 
a temperature dependence of all reaction rates, as well as preserving 
material frame invariance and equilibrium constrains.}

\cred{The novelties of this contribution also include a mixed-primal method based 
on a pressure-robust formulation for hyperelasticity. Our numerical scheme has been 
used to assess the influence of space and time
discretisation at different thermal states in three-dimensional 
domains. Comparisons were made in terms of local conduction velocity as well as 
onset and development of scroll wave dynamics as precursors of life threatening arrhythmias.
The numerical simulations demonstrated the suitability of the proposed model in reproducing 
key physiological features. In addition, we have observed model scalability adequate to conduct 
large scale computations.}

\cred{Our results, collected in Section~\ref{sec:results}, suggest that the new model 
develops higher nonlinearities and allows for more complex fibrillation dynamics 
when simulating classical S1-S2 stimulation protocols in anisotropic 
ventricular domains. For instance, the presence of cold regions in combination with our 
active strain model lead to an enhanced cardiac dispersion of repolarisation, which 
in turn results into more involved scroll wave dynamics.
Stimulation protocols (which represent the possible initiation of spiral waves 
and arrhythmic patterns from e.g. a fictitious ectopic focus)  
are greatly affected, and might even fail, under modified 
temperature conditions. For instance, allowing temperature gradients along 
or across the fibre direction can result in completely different activation patterns. 
A thorough computational assessment of these differences is therefore of 
key importance in determining experimental pacing mechanisms \citep{gizzi:2017a}.
We believe that the disruptions produced \emph{uniquely} by temperature gradients 
can be even more pronounced in the context of 
electromechanical simulations (as a consequence of the 
nonlinear coupling between the involved effects), and thus play a 
key role in the onset and development of arrhythmias. 
The set of preliminary tests presented in this paper 
highlights the importance of the proposed 
thermo-electro-mechanical coupling. Nevertheless, 
further investigations are necessary to determine other  
potential effects of the thermal coupling into the 
formation of local anchoring of spiral waves to material heterogeneities 
(pinning phenomena, \citealp{cherubini:2012}), 
their removal through low energy intra-cardiac defibrillators (unpinning protocols, see for instance \citealp{luther:2011})
and also the influence of the mechanochemical patterns in the 
induction and modulation of spatio-temporal alternans dynamics.}

\cred{General limitations of our study reside in that we adopt a simplified phenomenological model for both the 
thermo - electrophysiology and the excitation contraction coupling. Also, 
we have employed only idealised geometries in all our computations, 
but remark that a more dedicated personalisation could be incorporated once 
the following list of possible generalisations are in place. }

First, higher complexity in the electrophysiology and in the
contraction models should be \cred{included to improve 
the (at this point, still quite basic) structure of the coupling mechanisms. In particular, these 
extensions could lead to more
refined conclusions regarding the onset and control of arrhythmias and
fibrillation.}  Secondly, it is left to investigate whether
spatio-temporal variations of temperature have an effect, perhaps in
long term and operating theatre scenarios.
In perspective, the present study could serve in understanding and 
possibly controlling temperature-altered cardiac dynamics in patients 
subjected to whole-body hyperthermia. This is a medical procedure 
relevant for treating metastatic cancer and severe viral infections as e.g. HIV 
\citep{jha:2016,kinsht:2006}. 
\cred{Let us also remark that heat conduction in the short-scales we consider here (that is 
within one or two heart beats), can still be considered negligible. However, energy dissipation within an extended 
non-equilibrium thermodynamics framework could be an important improvement to our models. These 
extensions would incorporate a complete bio-heat formulation \citep{pennes,gizzi:2010},  which 
can account for the combined effects of heat generation from the 
heart muscle, as well as advection-diffusion of temperature due to vasculature and blood flow.} 

\cred{Another limitation of the present model is the phenomenological description of the intracellular calcium. 
The lack of precise calcium dynamics forces us to include an \emph{ad hoc} calcium-stretch coupling. More realistic models 
as the one in e.g. \citep{land:2017} account also for better action potential shape and morphology, inter- and intracellular calcium dynamics and potentially including multiscale thermo-mechanical features; they will be incorporated in our framework in a next stage. 
On the same lines, we also aim at incorporating microstructure-based bidomain formulations \citep{richardson:2011}, but specifically targeted for electromechanical couplings \citep{sharma:2018}.} 

In addition, we plan to apply the present model and computational 
methodology in the study of spatio-temporal alternans \citep{dupraz:2015}
as well as spiral pinning and unpinning phenomena~\citep{horning:2012, zhang:2018}. 
These supplementary studies would also contribute to further validate the 
proposed multiphysics framework against experimental evidence. 
In fact, the idealised ventricular domain embedded with myocardial fibres in 
rotational anisotropy that we used in Section~\ref{sec:ellipsoid} 
can be readily exploited towards the characterisation of the \cred{complex and}
unknown intramural dynamics, \cred{as soon as high-resolution imaging data
are incorporated into our computational framework.}
A further tuning of the material parameters 
using synchronised endocardial and epicardial optical mapping datasets 
will also be carried out. These studies are considered as a direct application of 
the very recent technology developed in \cite{christoph:2018} 
for the identification of electromechanical waves, and phase singularities in particular, through advanced imaging procedures.

The passive material properties of the
muscle have been considered independent of temperature, and a simple
constitutive relation could be embedded in the strain-stress law using
the results from \cite{templeton:1974}, or in the cell contraction
model following e.g. \cite{lu:2006}. 
\cred{Moreover, taking as an example what has been proposed for other biological scenarios such as 
vascular pathologies, we estimate that the concepts of 
time-dependent mechanobiological stability~\citep{cyron:2014} as well as 
 growth and remodelling~\citep{cyron:2016,cyron:2017b}, 
could be incorporated in the context of computational modelling of the heart. 
For instance, here we would}
consider distributed properties of collagen and muscular fibres, following for instance~\cite{pandolfi:2016}. 
Mechanoelectric feedback has been left out from this study (with the aim of isolating
the effects of the thermo-electric contribution in an
electromechanical context), so we could readily employ recent models
for stretch activated currents
\citep{quinn:2015,ward:2008}, or alternatively employ 
stress-assisted conductivity as in \cite{cherubini:2017}. Other
extensions include the use of geometrically detailed biventricular
meshes, more sophisticate boundary conditions (setting for instance
pressure-volume loops on the endocardium), and the presence of Purkinje networks~\citep{costabal:2016}
and/or fast conduction systems that could definitely have an impact on
the reentry dynamics. Goals in the longer term deal with optimal
control problems exploiting data assimilation techniques \citep{barone:2017} using imaging
tools and \emph{in silico} testing of novel defibrillation protocols~\citep{luther:2011}.

\bigskip 
\small 
\subsection*{Acknowledgments}
\cred{
This work has been partially supported by the  Engineering
and Physical Sciences Research Council (EPSRC) through the research
grant EP/R00207X/; by the Italian National Group
of Mathematical Physics (GNFM-INdAM); and by the International Center for
Relativistic Astrophysics Network (ICRANet). In addition, fruitful discussions with Daniel E. Hurtado (PUC), 
Bishnu Lamichhane (Newcastle), Francesc Levrero (Oxford), Simone Pezzuto (USI), Adrienne Propp (Oxford), and Simone Rossi (Duke), 
are gratefully acknowledged.   
Finally, we thank the constructive criticism of two anonymous referees whose suggestions 
lead to several improvements with respect to the initial version of the manuscript.}

\bibliographystyle{plainnat} 
\bibliography{BibFileRuizBaier-cicp}
\end{document}